\documentclass{pjarXive}
\usepackage[normalem]{ulem}
\usepackage{upgreek}
\usepackage{combelow}
\usepackage{comment}
\usepackage{cite}
\usepackage{tikz,graphicx}
\usepackage{amsmath}
\usepackage{color}
\usepackage[mathscr]{eucal}
\usepackage{amssymb}

\def\XXint#1#2#3{{\setbox0=\hbox{$#1{#2#3}{\int}$}
     \vcenter{\hbox{$#2#3$}}\kern-.5\wd0}}

\newcommand{\be}[1]{\begin{equation} \label{#1} }
\newcommand{\bea}[1]{\begin{eqnarray} \label{#1} }

\newcommand{\bfi}{\begin{figure}}
\newcommand{\efi}{\end{figure}} 
\newcommand{\ee}{\end{equation}}
\newcommand{\eea}{\end{eqnarray}}
\newcommand{\bib}{\bibitem}
\newcommand{\lbl}{\label}
\newcommand{\smT}{\mbox{\small\it T}}

\newcommand{\w}{{\omega}}

\newcommand{\vk}{{\bf k}}
\newcommand{\ve}{{\bf e}}
\newcommand{\va}{{\bf a}}
\newcommand{\vb}{{\bf b}}
\newcommand{\vc}{{\bf c}}
\newcommand{\vg}{{\bf g}}
\newcommand{\vch}{{\bf \hat{c}}}
\newcommand{\vh}{{\bf h}}
\newcommand{\vj}{{\bf j}}

\newcommand{\vv}{{\bf v}}

\newcommand{\vp}{{\bf p}}
\newcommand{\vm}{{\bf m}}
\newcommand{\vrp}{{\bf r}}

\newcommand{\vd}{{\bf d}}

\newcommand{\vH}{{\bf H}}
\newcommand{\vE}{{\bf E}}
\newcommand{\vM}{{\bf M}}
\newcommand{\vP}{{\bf P}}

\newcommand{\vA}{{\bf A}}
\newcommand{\vB}{{\bf B}}
\newcommand{\vF}{{\bf F}}
\newcommand{\vD}{{\bf D}}
\newcommand{\vJ}{{\bf J}}

\newcommand{\vS}{{\bf S}}

\newcommand{\dI}{\overline{\bf I}}

\newcommand{\dmu}{\overline{\mbox{\boldmath $\mu$}}}
\newcommand{\dep}{\overline{\mbox{\boldmath $\epsilon$}}}
\newcommand{\dnu}{\overline{\mbox{\boldmath $\nu$}}}
\newcommand{\dtau}{\overline{\mbox{\boldmath $\tau$}}}

\newcommand{\vcJ}{\mbox{\boldmath ${\mathcal J}$}}
\newcommand{\vcP}{\mbox{\boldmath ${\mathcal P}$}}
\newcommand{\vcM}{\mbox{\boldmath ${\mathcal M}$}}
\newcommand{\vcE}{\mbox{\boldmath ${\mathcal E}$}}

\newcommand{\vcB}{\mbox{\boldmath ${\mathcal B}$}}

\newcommand{\vnh}{{\bf \hat{n}}}
\newcommand{\lrho}{{\mbox{\large $\rho$}}}

\newcommand{\eps}{\epsilon}

\begin{document}
\setcounter{page}{1}
\pjheader{
}

\title[
]
{Classical Power and Energy Relations for Macroscopic Dipolar Continua Derived from the Microscopic Maxwell Equations}
\footnote{\it 
}  \footnote{\hskip-0.12in\, 
}
\footnote{\vspace{-10mm}\hskip-0.12in\textsuperscript{} 
}
\author{Arthur~D.~Yaghjian
}
\runningauthor{
}
%
%
%
\begin{abstract}
Positive semi-definite expressions for the time-domain macroscopic energy density in  passive, spatially nondispersive, dipolar continua are derived from the underlying microscopic {Maxwell} equations satisfied by classical models of discrete bound dipolar molecules or inclusions of the material or metamaterial continua.  The microscopic derivation reveals two distinct positive semi-definite macroscopic energy expressions, one that applies to diamagnetic continua (induced magnetic dipole moments) and another that applies to paramagnetic continua (alignment of ``permanent" magnetic dipole moments), which includes ferro(i)magnetic and antiferromagnetic materials.  The diamagnetic dipoles are ``unconditionally passive" in that their Amperian (circulating electric current) magnetic dipole moments are zero in the absence of applied fields.  The analysis of paramagnetic continua, whose magnetization is caused by the alignment of randomly oriented ``permanent" Amperian magnetic dipole moments that dominate any induced diamagnetic magnetization, is greatly simplified by first proving that the microscopic power equations for rotating ``permanent" Amperian magnetic dipoles  (which are shown to not satisfy unconditional passivity) reduce effectively to the same power equations obeyed by rotating unconditionally passive magnetic-charge magnetic dipoles.  The difference between the macroscopic paramagnetic and diamagnetic energy expressions is equal to a ``hidden energy" that parallels the hidden momentum often attributed to  Amperian magnetic dipoles.  The microscopic derivation reveals that this hidden energy is drawn from the reservoir of inductive energy in the initial paramagnetic microscopic Amperian magnetic dipole moments.  The macroscopic, positive semi-definite, time-domain energy expressions are applied to lossless bianisotropic media to determine the inequalities obeyed by the frequency-domain bianisotropic constitutive parameters.  Subtleties associated with the causality as well as the group and energy-transport velocities for diamagnetic media are discussed in view of the diamagnetic inequalities.
\end{abstract}
%
%
\section{Introduction}
\begin{figure}[ht]
\begin{center}
\includegraphics[width =4.0in]{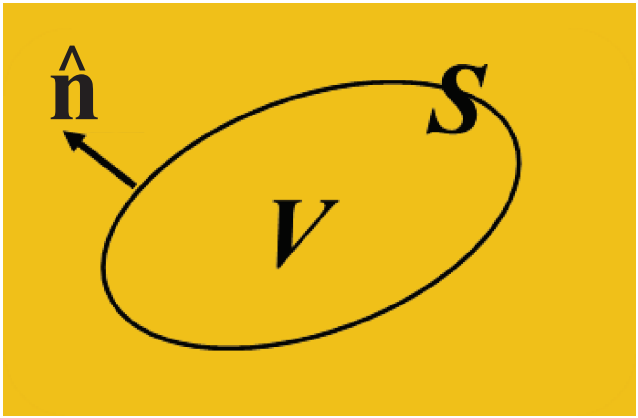}
\end{center}
\caption{\label{fig1}Volume $V$ with surface $S$ in dipolar continua with macroscopic polarization densities $\vP$ and $\vM$.}
\end{figure}

As a way of introducing the motivation for and purpose of this paper, begin with the familiar Maxwell macroscopic equations that hold in dipolar continua {(without current $\vJ$)} to obtain the macroscopic {Poynting} theorem as\footnote{{In reading Poynting's original 1884 paper, one finds that he wrote his theorem in the form (using modern notation and SI units with displacement current written as $\partial \vD/\partial t$ instead of $\eps\partial \vE/\partial t$ {and $\mu\partial\vH/\partial t$ written as $\partial\vB/\partial t$})  
\be{footPoy}
\int\limits_V \left\{\frac{\partial \vD}{\partial t}\cdot \vE_{\rm v} + \frac{\partial \vB}{\partial t}\cdot \vH + \vJ\cdot\vE_{\rm v} +\left[\left(\vJ +\frac{\partial \vD}{\partial t}\right)\times\vB\right]\cdot\vv\right\} dV = -\int\limits_S \vnh\cdot(\vE\times\vH)dS
\ee
where $\vE_{\rm v} = \vE+\vv\times\vB$ {and $\vv$ is the velocity of the ``matter"} \cite[eq. (7)]{Poynting}.  This is not too surprising since Poynting references Maxwell's Treatise in which Maxwell wrote some of his final equations in terms of $\vE +\vv\times\vB$ \cite{Maxwell}, \cite{Yaghjian-Reflection}.}  {The $\vv$ terms in (\ref{footPoy}) cancel to leave the familiar form of ``Poynting's theorem".}}\\[-3mm]
\be{I1}
-\int\limits_S \vnh\cdot[\vE(\vrp,t)\times\vH(\vrp,t)] dS =\int\limits_V \left[\frac{\partial\vD(\vrp,t)}{\partial t}\cdot\vE(\vrp,t)
+ \frac{\partial\vB(\vrp,t)}{\partial t}\cdot\vH(\vrp,t) \right] dV
\ee
\mbox{}\\[-4mm]
with\\[-4mm]
\be{I1'}
\vD = \eps_0\vE +\vP\,,\;\;\;\; \vB = \mu_0(\vH+\vM)
\ee
where ($\vE,\vD$) and ($\vB,\vH$) are the macroscopic electric and magnetic fields, respectively, $\vP$ and $\vM$ are the macroscopic electric and magnetic dipolarization densities, respectively, and  $\eps_0$ and $\mu_0$ are the free-space permittivity and permeability, respectively.  (The term ``macroscopic" is used throughout to designate electromagnetic fields and sources averaged over macroscopic volumes that contain a large number of discrete dipolar molecules or inclusions while having dimensions much smaller than the minimum temporal and spatial wavelengths in free-space and in the continuum.  An electromagnetic ``dipolar continuum" or just ``continuum" is used throughout to refer to a medium in which fields and sources obey the traditional Maxwell dipolarization equations; see (\ref{12}) or (\ref{12'}).  We will later find it useful to distinguish between ``ideal" and ``macroscopic" continua.)   The unit normal $\vnh$ to the closed surface $S$ points out of its volume $V$ so that the left-hand side (and thus the right-hand side) of (\ref{I1}) is {(as will be proven later)} equal to the instantaneous electromagnetic power flow entering the volume $V$; {see Fig. \ref{fig1}}.  Integrating (\ref{I1}) from time $t_0$ \textit{when the macroscopic fields are zero} to the present time $t$ gives a macroscopic electromagnetic energy density  on the right-hand side of (\ref{I1}) equal to
\be{I2}
\int\limits_{t_0}^t \left[\frac{\partial\vD(\vrp,t')}{\partial t'}\cdot\vE(\vrp,t')
+ \frac{\partial\vB(\vrp,t')}{\partial t'}\cdot\vH(\vrp,t') \right]  dt'\,.
\ee
\par
If it is assumed that the macroscopic electromagnetic energy in a passive polarized continuum is at least as great as the energy of the same $\vE$ and $\vH$ fields produced in a vacuum, then this vacuum energy {density} is $(\eps_0|\vE|^2 +\mu_0|\vH|^2)/2$ and (\ref{I2}) yields the inequality
\be{I52''}
\int\limits_{t_0}^t \left[\frac{\partial\vD(\vrp,t')}{\partial t'}\cdot\vE(\vrp,t')
+ \frac{\partial\vB(\vrp,t')}{\partial t'}\cdot\vH(\vrp,t') \right] dt' \ge \frac{1}{2}\left[\eps_0 |\vE(\vrp,t)|^2 +\mu_0|\vH(\vrp,t)|^2\right]
\ee
or simply
\be{I3}
\int\limits_{t_0}^t \left[\frac{\partial\vP(\vrp,t')}{\partial t'}\cdot\vE(\vrp,t')
+ \mu_0\frac{\partial\vM(\vrp,t')}{\partial t'}\cdot\vH(\vrp,t') \right] dt' \ge 0\,.
\ee
\par
If, on the other hand, one takes the view that all known magnetization is produced by Amperian magnetic dipoles (circulating electric current) and thus $\vE$ and $\vB$ are the primary fields, then the vacuum energy {density} is $(\eps_0|\vE|^2 +|\vB|^2/\mu_0)/2$ and (\ref{I2}) yields instead of (\ref{I52''}) and (\ref{I3})
\be{I52'''}
\int\limits_{t_0}^t \left[\frac{\partial\vD(\vrp,t')}{\partial t'}\cdot\vE(\vrp,t')
+ \frac{\partial\vB(\vrp,t')}{\partial t'}\cdot\vH(\vrp,t') \right] dt' \ge \frac{1}{2}\left[\eps_0 |\vE(\vrp,t)|^2 +\frac{1}{\mu_0}|\vB(\vrp,t)|^2\right]
\ee
or simply
\be{I4}
\int\limits_{t_0}^t \left[\frac{\partial\vP(\vrp,t')}{\partial t'}\cdot\vE(\vrp,t')
- \frac{\partial\vB(\vrp,t')}{\partial t'}\cdot\vM(\vrp,t') \right] dt' \ge 0\,.
\ee
%
\par
Although it is often argued that (\ref{I3}) is the correct inequality for passive continua \cite{Glasgow, Mansuripur2013, Welters},  \cite[p. 81]{F&M}, in fact, neither one of the inequalities in (\ref{I3}) or (\ref{I4}) are universally valid because the simple hypothetical example of a passive material with constitutive relations $\vD = \eps\vE$ and $\vB =\mu\vH$ with real constant permittivity $\eps$ and permeability $\mu$ over the operational bandwidth reveals from (\ref{I3}) that
\be{I5}
\eps \ge \eps_0\,,\;\;\;\; \mu\ge\mu_0
\ee
and from (\ref{I4}) that
\be{I6}
\eps \ge \eps_0\,,\;\;\;\; 0\le\mu\le\mu_0\,.
\ee
This would suggest that (\ref{I3}) and (\ref{I4}) may be valid for {paramagnetic} and diamagnetic materials, respectively.  Indeed, the main purpose of this paper is to show that, under quite general sufficient conditions \textit{and without assuming any particular constitutive parameters},  (\ref{I3}) and (\ref{I4}) apply to {paramagnetic} and diamagnetic dipolar continua, respectively.
\par
To prove these positive semi-definite time-domain inequalities in (\ref{I3}) and (\ref{I4}), the macroscopic Maxwell equations and Poynting theorem are systematically related to the more fundamental microscopic Maxwell equations and Poynting theorem from which their macroscopic counterparts are derived.  Diamagnetic material \textit{is defined} simply as material with Amperian magnetization produced by molecules or inclusions that have no {permanent} magnetic dipole moments, only induced magnetic dipole moments.  {Paramagnetic material, a term that includes ferro(i)magnetic and antiferromagnetic material \cite[chs. 11 and 12]{Kittel}, \textit{is defined} simply as material with magnetization produced by the alignment of ``permanent" (sometimes called ``intrinsic") Amperian magnetic dipole moments of molecules or inclusions that dominate any induced diamagnetic magnetization.\footnote{\lbl{foot0}These fundamental definitions of diamagnetism and paramagnetism differ from the more common definitions as materials (or metamaterials) with negative and positive susceptibilities, respectively \cite{Kittel}.  The more fundamental definitions allow for the possibility of both diamagnetic and paramagnetic susceptibilities to be positive or negative at higher frequencies where resonances may occur.}  Notably, the detailed microscopic derivation using classical models of electric and magnetic dipoles reveals that the difference between the integrals of the power densities in (\ref{I3}) and (\ref{I4}), what can be called the macroscopic hidden energy, is drawn from the reservoir of inductive energy in the permanent (initial paramagnetic) microscopic Amperian magnetic dipoles.  (The term ``permanent" does not imply that the value of the initial current and magnitude of the  magnetic dipole moment cannot change slightly as the magnetic dipole rotates in an external field.)
\par
Besides the Introduction and Conclusion, the paper contains six other main sections:
\par
In Section \ref{ME}, the mathematically rigorous vector-scalar potential solution is given to Maxwell's differential equations for microscopic electric current and charge density. From these equations, it is {established} that the microscopic Poynting vector integrated over a closed surface in free space equals the total instantaneous power crossing that free-space surface. It is further shown that, in passive materials, the energy supplied by the fields to the microscopic charge carriers from the time the external fields are first applied is always positive semi-definite (nonnegative).
\par
In Section \ref{DC}, Maxwell's continuum equations are derived from the microscopic equations of Section \ref{ME} for an ideal continuum, defined by infinitesimal subvolumes of electric and magnetic dipolarization that are continuous, and a straightforward proof is given that these same Maxwell equations describe the macroscopic dipolarization obtained from averaging the sources and fields of classical models of discrete electric and magnetic dipoles.  It is shown that a requirement for the equivalence of the ideal and macroscopic continua equations is that the surfaces of the macroscopic averaging volumes lie in free space and not cut through the discrete dipoles.
\par
In Section \ref{PTDC}, the conditions are found ({namely} that the continuum is effectively spatially nondispersive) for the macroscopic continuum Poynting vector integrated over a closed surface to equal the instantaneous power flow across that closed surface.
\par
Section \ref{ER} begins with a derivation of expressions for macroscopic continua energy densities in terms of the microscopic energy densities.  It is determined that, under sufficient conditions that are designated by the term ``unconditionally passive", one of these energy-density expressions remains positive semi-definite for diamagnetic media, and a second different energy-density expression remains positive semi-definite for paramagnetic media.
\par
In Section \ref{RER}, the important positive semi-definite energy-density expressions for macroscopic continua are summarized for paramagnetic and diamagnetic continua and it is shown that the difference between these two energy densities equals a ``hidden energy" extracted from the energy in the ``permanent" Amperian microscopic magnetic dipole moments that align to produce the paramagnetic dipolarization.
\par
In Section  \ref{BIAN}, the positive semi-definite time-domain inequalities of Section \ref{RER} are applied to {paramagnetic} and diamagnetic bianisotropic continua to derive inequalities satisfied by lossless frequency-domain bianisotropic constitutive parameters.   Lastly, subtleties associated with diamagnetic continua are discussed, namely, the causality of diamagnetic permeability as well as the group and energy-transport velocities in lossless diamagnetic media.
\par
We employ a classical analysis of macroscopic dipolar continua in which no attempt is made to determine the detailed quantum nature of the polarizations and fields of atoms and molecules.  It is assumed, as in most texts on the electromagnetic fields and equations of polarized media, that since most materials below optical frequencies and dipolar metamaterials well below bandgap frequencies are adequately described by the classical Maxwell macroscopic equations for a dipolar continuum, the microscopic dipoles producing the macroscopic dipolarization can be adequately modeled pragmatically by classical electric-charge electric dipoles and Amperian-current magnetic dipoles, irrespective of their actual quantum origin.
\par
There are several reasons for determining nonnegative macroscopic energies from the microscopic {Maxwell} equations for classical diamagnetic and paramagnetic dipolar continua.  First, it is one of the remaining unresolved problems in classical electromagnetic theory even though the problem is relatively easy to state: given a volume of a macroscopic continuum satisfying Maxwell's dipolar equations and illuminated by external fields, are there macroscopic polarization energy densities that never become negative for all times after an initial time when the macroscopic fields and polarizations are zero. 
Nonnegative energy expressions provide a means for determining realistic physical limitations, such as the upper bounds on the bandwidth and gain of antennas or the lower bounds on antenna quality factors.  Valuable inequalities satisfied by bulk constitutive parameters as well as by the group and energy transport velocities in materials and metamaterials can be obtained from nonnegative macroscopic energy expressions.  Also, knowing the sufficient conditions for the validity of the nonnegative macroscopic energy expressions lends insight into the development and utilization of materials and metamaterials that may not be subject to the restrictions imposed by these expressions.
\section{\label{ME}{MAXWELL'S EQUATIONS FOR MICROSCOPIC ELECTRIC CHARGE AND CURRENT}}
We assume that we are dealing with macroscopic continua whose molecules or inclusions can be modeled by {classical} microscopic electric charge and current whose fields can be adequately described by the following Maxwell differential equations in SI (mksA) units
\begin{subequations}
\label{1}
\be{1a}
\nabla\times\ve(\vrp,t) +\frac{\partial \vb(\vrp,t)}{\partial t} =0
\ee
\be{1b}
\frac{1}{\mu_0}\nabla\times\vb(\vrp,t) -\eps_0\frac{\partial \ve(\vrp,t)}{\partial t} = \vj(\vrp,t)
\ee
\be{1c}
\nabla\cdot\vb(\vrp,t) =0
\ee
\be{1d}
\eps_0\nabla\cdot\ve(\vrp,t) = \varrho(\vrp,t)
\ee
\end{subequations}
where $\ve(\vrp,t)$ and $\vb(\vrp,t)$ are the primary microscopic electric and magnetic fields at the position $\vrp$ and time $t$, $\varrho(\vrp,t)$ and $\vj(\vrp,t)$ are the microscopic electric charge and current densities, and $\eps_0$ and $\mu_0$ are the free-space permittivity and permeability, respectively.  The primary electric and magnetic fields, $\ve(\vrp,t)$ and $\vb(\vrp,t)$, are defined by the instantaneous Lorentz force, $\vF(\vrp,t) = q[\ve(\vrp,t)+\vv\times\vb(\vrp,t)]$, that they exert on a point test charge $q$ moving with velocity $\vv$ at the time $t$ and position $\vrp$.  (The point test charge $q$ is assumed small enough that it does not disturb the sources of the electric and magnetic fields being measured.  The sources of $\ve(\vrp,t)$ and $\vb(\vrp,t)$ are external to $q$, that is, they do not include the fields at $(\vrp,t)$ produced by the moving test charge $q$.  Also, any radiation reaction force on $q$ is assumed negligible, as it will be if the acceleration of the charge is negligible or if the charge $q$ is sufficiently small since the radiation force is proportional to the square of $q$ \cite{Yaghjian-LEbook}.)  Note that since there are no microscopic polarization densities ($\vp$ and $\vm$) in (\ref{1}), it follows that the microscopic electric displacement vector $\vd$ is given by $\vd =\eps_0\ve$ and the microscopic secondary magnetic field $\vh$ is given by $\vh =\vb/\mu_0$.\footnote{\lbl{foot1} For metamaterial inclusions with microscopic polarization densities $\vp$ and $\vm$,  the current $\vj$  { in (\ref{1})} would include $\partial \vp/\partial t {+ \nabla\times\vm}$, and the charge $\varrho$ would include $-\nabla\cdot\vp$.  Thus, the main results of this paper are applicable to these metamaterials as well within the bandwidths that they behave as dipolar continua \cite{YAS1,YAS2}, provided either paramagnetism or diamagnetism dominates the {macroscopic} magnetization produced by the inclusions.  For example, metal-dielectric inclusions in metamaterial-array continua would be described by the electric-dipole/diamagnetic energy expressions.}
\par
The rigorous vector-scalar potential solution to Maxwell's equations (\ref{1}) in the frequency domain (with $e^{-i\w t}$ time dependence) is given by \cite[sec. 2.3.6]{H&Y}
\begin{subequations}
\label{1'}
\be{1'a}
\vb_\w(\vrp) = \nabla\times \va_\w(\vrp)
\ee
\be{1'b}
\ve_\w(\vrp) = i\w\va_\w(\vrp) -\nabla \psi_\w(\vrp)
\ee
\end{subequations}
with the vector and scalar potentials determined from the expressions
\begin{subequations}
\label{1''}
\be{1''a}
\va_\w(\vrp) = \mu_0\lim_{\delta\to0}\int\limits_{V-V_\delta}\vj_\w(\vrp')G(|\vrp-\vrp'|) dV'
\ee
\be{1''b}
\psi_\w(\vrp) =  \frac{1}{\eps_0}\lim_{\delta\to0}\int\limits_{V-V_\delta}\varrho_\w(\vrp')G(|\vrp-\vrp'|) dV'
\ee
and the scalar Green's function given by
\be{1''c}
G(|\vrp-\vrp'|) = \frac{e^{ik_0|\vrp-\vrp'|}}{4\pi|\vrp-\vrp'|},\;\;\;k_0 = \w\sqrt{\mu_0\eps_0}\,.
\ee
\end{subequations}
The limits before the integrals of (\ref{1''}) are rigorously required to isolate the singularity in the Green's function at $\vrp' =\vrp$ from the volume integration by a ``principal volume" $V_\delta$ (with maximum dimension $\delta$) to allow rigorous differentiation of the potentials to obtain the fields in (\ref{1'}) \cite[sec. 2.3.6]{H&Y},  \cite{Yaghjian-DGF}.
\subsection{\label{PTcc}{Poynting's theorem for microscopic electric charge and current}}
Let the microscopic charge and current be confined to a region of finite extent so that they can be enclosed by a volume $V$ whose surface $S$ lies in free space.  Then the instantaneous power $P_{je}(t)$ supplied by the fields to the charge-current within $V$ is simply 
\be{2}
P_{je}(t) = \int\limits_V \vj(\vrp,t)\cdot\ve(\vrp,t) dV\,.
\ee
This is proven from the Lorentz force on a volume charge density $\varrho_{\rm v}$ moving with velocity $\vv$, namely $d\vF = \varrho_{\rm v}(\ve +\vv\times\vb)dV$ with $dP = \vv\cdot d\vF = \varrho_{\rm v}\vv\cdot\ve dV = \vj\cdot\ve dV$.\footnote{\lbl{foot2} The Lorentz force on $\varrho dV$ is exerted by the fields of all sources except the fields of $\varrho dV$ itself.  However, the self-fields of an infinitesimal volume element of continuous volume density of charge and current approach zero as $dV \to 0$ and thus $\ve$ and $\vb$ in this volume-element Lorentz force are the total electric and magnetic fields in Maxwell's equations (\ref{1}).}  (Thus, the magnetic field does not impart energy to the charge carriers.)   With the help of (\ref{1a}) and (\ref{1b}), the power in (\ref{2}) can be recast in the form of the microscopic {Poynting} theorem
\be{3}
P_{je}(t) = \int\limits_V \vj\cdot\ve dV = -\int\limits_V \left[\nabla\cdot(\ve\times\vb/\mu_0) +\frac{1}{2}\frac{\partial}{\partial t}\left(\eps_0|\ve|^2 +|\vb|^2/\mu_0\right)\right] dV
\ee
or by means of the divergence theorem as
\be{4}
\int\limits_S \vnh\cdot(\ve\times\vb/\mu_0) dS =   -\int\limits_V \vj\cdot\ve dV -  \frac{1}{2}\frac{d}{dt}\int\limits_V\left(\eps_0|\ve|^2 +|\vb|^2/\mu_0\right) dV
\ee
with $\vnh$ denoting the unit normal to $S$ pointing out of the volume $V$.
\par
As any one of an infinite number of test examples to which (\ref{4}) can be applied, suppose that the volume $V$ contains a charged capacitor $C$ that is shorted at $t=0$ through an inductor $L$.  Even if the capacitor, the inductor, and the wires are {lossless conductors}, this LC circuit will not oscillate forever because it will radiate.  At the time $t= \smT$, assume that the radiation has practically reduced the oscillations of the RL circuit to zero and that the surface $S$ is chosen far enough away that the radiated fields have not reached $S$ at the time $\smT$.  Then (\ref{4}) becomes after integrating over time from $0^-$ to $\smT$
\be{5}
\frac{1}{2}\int\limits_V\eps_0|\ve(\vrp,0^-)|^2  dV = \int\limits_{0^-}^{T}\int\limits_V \vj(\vrp,t)\cdot\ve(\vrp,t) dV dt + \frac{1}{2}\int\limits_V\left[\eps_0|\ve(\vrp,\smT)|^2 +|\vb(\vrp,\smT)|^2/\mu_0\right] dV
\ee
where $0^-$ denotes the time just before $t=0$.
The left-hand side of (\ref{5}) is equal to the electrostatic energy stored initially by the capacitor, a result that is rigorously proven from the electrostatic equations under a quasi-static charging of the capacitor plates.  As explained above, the first integral on the right-hand side of (\ref{5}) is the energy supplied by the fields to the charge carriers of the current. (It is zero for a stationary {lossless conductor and greater than zero for a lossy conductor.})  Therefore, the second integral on the right-hand side of (\ref{5}) is an additional energy required by conservation of energy.  {\em In other words, $\int_V[\eps_0|\ve(\vrp,t)|^2 +|\vb(\vrp,t)|^2/\mu_0] dV /2$ is the free-space energy stored in the electromagnetic fields in the volume $V$ not only for statics but also for time dependent fields at each instant of time $t$.}
With this established, we see that the right hand side of (\ref{4}), and thus the left-hand side of (\ref{4}), is the time rate of change of the total energy leaving the volume $V$.  In particular, {\em the integral of the time-dependent microscopic Poynting vector, that is, $\vnh\cdot[\ve(\vrp,t)\times\vb(\vrp,t)/\mu_0]$, over a closed surface $S$ lying in free space is the total instantaneous electromagnetic power leaving that closed surface}.  Alternatively, the total instantaneous electromagnetic power $P(t)$ entering a closed surface $S$ that lies in free space is given by
\bea{5'}
P(t) = -\int\limits_S \vnh\cdot[\ve(\vrp,t)\times\vb(\vrp,t)/\mu_0] dS \hspace{51.5mm}\nonumber\\ =   \int\limits_V \vj(\vrp,t)\cdot\ve(\vrp,t) dV  + \frac{1}{2}\frac{d}{dt}\int\limits_V\left[\eps_0|\ve(\vrp,t)|^2 +|\vb(\vrp,t)|^2/\mu_0\right] dV\,.
\eea
This result is true whether or not the microscopic charge-current inside $S$ forms polarized material, provided $S$ lies in free space outside the material.  However, this result does not imply that $\vnh\cdot(\ve\times\vb/\mu_0)$ is necessarily the instantaneous free-space power flow per unit area in the direction of $\vnh$.  In other words, the integral of $\vnh\cdot(\ve\times\vb/\mu_0)$ over an open free-space surface does not necessarily equal the electromagnetic power that flows across that open surface.
\subsubsection{\label{Energy}{Energy supplied to the charge carriers}}
The instantaneous power supplied by the electromagnetic fields  to the charge carriers (usually electrons and positively charged nuclei)  
 producing the microscopic charge and current ($\varrho,\vj$) that generate the fields $\ve$ and $\vb$ under consideration in $V$ is given in (\ref{2}).   Expressing $P_{je}(t)$ as the time derivative of an energy $W_{je}(t)$, that is,  $P_{je}(t)=dW_{je}(t)/dt$, and integrating the power in (\ref{2}) from an initial time $t_0$ in the past to the present time $t$, we obtain
\be{4W}
W_{je}(t)-W_{je}(t_0) =\int\limits_{t_0}^t \int\limits_V \vj(\vrp,t')\cdot\ve(\vrp,t') dV dt'\,.
\ee
This is the energy supplied to (work done on) the charge carriers of the current $\vj$ in $V$ by the electric field $\ve$ produced by the charge carriers during the time interval $t-t_0$.  \textit{Assume that the applied electromagnetic fields are zero until after the time $t_0$ so that the electric field does no work on the charge carriers until after $t_0$.}  If the charge-current of the dipoles is produced by bound charge carriers (charge-current comprising electric and magnetic dipoles with negligible translational displacement and negligible changes in their number per unit volume), then this work, $W_{je}(t)-W_{je}(t_0)$, done on the bound charge carriers equals the change in kinetic energy of the charge carriers supplied by $\ve$ plus any change in kinetic energy of the charge carriers caused by forces other than the $\ve$-field force on the charge carriers.\footnote{\lbl{foot4}Each differential element of charge $\varrho dV$ could radiate energy and thus experience an irreversible self-radiation-reaction energy as well as reversible self-radiation-reaction (often called the ``Schott acceleration energy") and self-electromagnetic-momentum energies \cite{Yaghjian-LEbook}.  However, since all these self-energies are proportional to the square of the charge, that is, $(\varrho dV)^2$, they are a higher-order differential than $dV$ and thus they can be ignored in the volume integral of (\ref{4W}).  (This does not imply that the integrated self-force/momentum and self-power/energy for a fixed amount of charge moving as a relativistically rigid charged particle is negligible \cite{Yaghjian-LEbook}.)} These other forces are fundamentally electromagnetic forces from the fields of the host material, namely from the fields of the atomic and molecular charges other than the charge carriers that produce $\ve$ and $\vb$.  They are the forces, for example, involved in the conversion of the kinetic energy of the charge carriers into heat (out-of-band energy), or into the kinetic-potential-heat energy of the springs in a compressed-spring model of electric dipoles, or into the kinetic-potential-heat energy of expandable conducting wires in a wire-loop model of Amperian magnetic dipoles. 
\par
If the macroscopic continuum is passive in that (i) there is negligible initial kinetic energy of the bound charge carriers within the operational bandwidth (that is, negligible in-band initial kinetic energy) that can be decreased, (ii) there is negligible in-band net energy transfer from the host material to the charge carriers (that is, in-band energy cannot travel through the host material, except by means of $\ve$ and $\vb$, from one region of the continuum to the other), and (iii) there are no auxiliary sources  (such as chemical reactions changing the $\ve$ and $\vb$ fields so as to add to the energy of the charge carriers, or initial kinetic-potential energy stored in the springs of a compressed-spring model of electric dipoles, or in expandable conducting wires of ``permanent" Amperian magnetic dipoles) \textit{that can release energy} upon excitation by applied fields to serve as an active source of internal energy increasing the energy of the charge carriers, then $W_{je}(t)-W_{je}(t_0)\ge0$.  In other words, if there is no initial in-band mechanical energy, and no internal active sources of energy, and no nonlocal transfer of in-band mechanical energy, then the material is passive such that $W_{je}(t)-W_{je}(t_0)\ge0$.   Choosing the arbitrary, constant initial energy $W_{je}(t_0)$ equal to zero, we have
\be{5W}
W_{je}(t) =\int\limits_{t_0}^t \int\limits_V \vj(\vrp,t')\cdot\ve(\vrp,t') dV dt' \ge 0
\ee
for bound charge carriers in {\textit{passive} material that can be lossy as well as lossless, linear or nonlinear.  Equation (\ref{5W}) can be taken as the mathematical definition of \textit{passivity}.  A more restrictive ``unconditional passivity" will be defined later in Section \ref{ER}.}
\par
The current $\vj(\vrp,t)$ in (\ref{5W}) can include conduction current (in addition to the current produced by the local motion of the bound charges) if the changes in the energy transferred by the drift velocity of the conduction charges are negligible compared with the heat energy produced by the conduction current (as is normally the case).
\section{\label{DC}{MAXWELL'S EQUATIONS FOR DIPOLAR CONTINUA}}
In order to explain how the microscopic electric and magnetic fields of discrete electric and magnetic dipoles should be averaged to obtain the traditional macroscopic {Maxwell} equations for dipolar media, it is helpful to first determine the ideal continuum Maxwell equations directly without assuming the existence of discrete dipoles.  This direct continuum approach for deriving Maxwell's dipolarization equations originates with Maxwell himself \cite{Maxwell, Yaghjian-Reflection}. 
\subsection{\label{IdealContinuum}{Maxwell's equations for ideal dipolar continua}}
The essence of the derivation of the ideal continuum {Maxwell} equations is to first determine the vector and scalar potentials in free space outside the differential volume elements with dipole moments and then to mathematically define these potentials within the polarization source regions by simply applying the same mathematical expressions within the source regions.  These {Maxwell} fields defined mathematically within the polarization source regions are then related to fields that can be measured in free-space cavities formed by removing an infinitesimal volume of polarized material without altering the remaining polarization. 
\par
Specifically, the vector and scalar potentials in free space \textit{outside} a distribution of differential volume elements of electric and magnetic dipolarization, {$\vP_\w dV$ and $\vM_\w dV$}, are given in the frequency domain (with $e^{-i\w t}$ time dependence) by  \cite[secs. 9.2--9.3]{Jackson}
\begin{subequations}
\label{6}
\be{6a}
\va_\w(\vrp) = \mu_0\int\limits_V\left[-i\w\vP_\w(\vrp')G +\vM_\w(\vrp')\times\nabla'G\right] dV' 
= \mu_0\int\limits_V\left[-i\w\vP_\w(\vrp') +\nabla'\times\vM_\w(\vrp')\right]G dV'
\ee
\be{6b}
\psi_\w(\vrp) = \frac{1}{\eps_0}\int\limits_V\vP_\w(\vrp') \cdot\nabla'G dV' 
= -\frac{1}{\eps_0}\int\limits_V\nabla'\cdot\vP_\w(\vrp')G dV'
\ee
\end{subequations}
where $G$ is given in (\ref{1''c}) and the volume-element contributions are integrated over a volume $V$ whose surface $S$ lies in free space so that it encloses all possible equivalent electric surface charge $-\vnh'\cdot\vP_\w$ and equivalent electric surface current $\vM_\w\times\vnh'$ (surface delta functions in $-\nabla'\cdot\vP_\w$ and $\nabla'\times\vM_\w$).
These equations can be rigorously derived from (\ref{1}) and the definition of infinitesimal dipole moments determined by electric charge separation and circulating electric current (Amperian magnetic dipoles) outside the sources ($\vrp \neq \vrp'$) of these dipole moments, {$\vP_\w dV$ and $\vM_\w dV$}.  We can express $\vP_\w$ and $\vM_\w$ in terms of bound charge and current densities $\varrho_{\w b}$ and $\vj_{\w b}$ as 
\begin{subequations}
\label{6'}
\be{6'a}
\vP_\w(\vrp)  =\lim_{\Delta V\to0}\frac{1}{\Delta V}\int\limits_{\Delta V} \varrho_{\w b}(\vrp+\vrp')\vrp' dV'
\ee
\be{6'b}
\vM_\w(\vrp) = \lim_{\Delta V\to0}\frac{1}{2\Delta V} \int\limits_{\Delta V} \vrp'\times\vj_{\w b}(\vrp+\vrp') dV'
\ee
\end{subequations}
with the position vector $\vrp$ inside $\Delta V$, whose surface $\Delta S$ does not intersect any of the charge and current densities that produce the electric and magnetic dipolarization, and $\int_{\Delta V} \varrho_{\w b}(\vrp) dV =0$.
The magnetic and electric fields in (\ref{1}) \textit{outside this dipolarization} are given in the frequency domain as in (\ref{1'})
\begin{subequations}
\label{7}
\be{7a}
\vb_\w(\vrp) = \nabla\times \va_\w(\vrp)
\ee
\be{7b}
\ve_\w(\vrp) = i\w\va_\w(\vrp) -\nabla \psi_\w(\vrp)\,.
\ee
\end{subequations}
\par
Next let us simply define mathematically the vector and scalar potentials inside as well as outside the sources $\vP_\w$ and $\vM_\w$ by the same integrals as in (\ref{6}), namely
\begin{subequations}
\label{8}
\be{8a}
\vA_\w(\vrp) = \mu_0\lim_{\delta\to0}\int\limits_{V-V_\delta}\left[-i\w\vP_\w(\vrp') +\nabla'\times\vM_\w(\vrp')\right]G dV'
\ee
\be{8b}
\Psi_\w(\vrp) =  -\frac{1}{\eps_0}\lim_{\delta\to0}\int\limits_{V-V_\delta}\nabla'\cdot\vP_\w(\vrp')G dV'
\ee
\end{subequations}
with the mathematically defined electric and magnetic fields given from (\ref{7}) as{\cite{Chew}}
\begin{subequations}
\label{9}
\be{9a}
\vB_\w(\vrp) = \nabla\times \vA_\w(\vrp)
\ee
\be{9b}
\vE_\w(\vrp) = i\w\vA_\w(\vrp) -\nabla \Psi_\w(\vrp)
\ee
\end{subequations}
where we have now used the symbols $\vB_\w$, $\vE_\w$, $\vA_\w$, and $\Psi_\w$ rather than $\vb_\w$, $\ve_\w$, $\va_\w$, and $\psi_\w$ because $\vB_\w$, $\vE_\w$, $\vA_\w$ and $\Psi_\w$ defined mathematically by (\ref{8})--(\ref{9}) are not necessarily equal to the microscopic fields $\vb_\w$, $\ve_\w$,  $\va_\w$, and $\psi_\w$ in the source regions of $\vP_\w$ and $\vM_\w$.  The principal-volume limits are reinstated in the integrals of (\ref{8}) because, unlike in the integrals of (\ref{6}), now $\vrp$ can lie in the source regions of $\vP_\w$ and $\vM_\w$ and thus the singularity of the Green's function $G(|\vrp-\vrp'|)$ at $\vrp =\vrp'$ has to be excluded.  Nonetheless, the principal value limits must be retained in (\ref{8}) to get the correct values of the fields (and their spatial derivatives) in the source regions in terms of $\vP_\w$ and $\vM_\w$ by means of the differentiations in (\ref{9}) and subsequent use of integral identities \cite{Yaghjian-DGF}.
\par
Comparison of (\ref{8})--(\ref{9}) with (\ref{1'})--(\ref{1''}), which satisfy the frequency-domain version of the Maxwell equations in (\ref{1}), reveals that the fields and polarizations of the ideal continuum also satisfy the frequency-domain version of the Maxwell equations in (\ref{1}) but with $\vE_\w$ and $\vB_\w$ replacing $\ve_\w$ and $\vb_\w$, respectively, and with $-i\w\vP_\w +\nabla\times\vM_\w$ and $-\nabla\cdot\vP_\w$ replacing $\vj_\w$ and $\varrho_\w$, respectively.  Consequently, the ideal continuum dipolarization fields satisfy the {Maxwell} equations
\begin{subequations}
\label{10}
\be{10a}
\nabla\times\vE(\vrp,t) +\frac{\partial \vB(\vrp,t)}{\partial t} =0
\ee
\be{10b}
\frac{1}{\mu_0}\nabla\times\vB(\vrp,t) -\eps_0\frac{\partial \vE(\vrp,t)}{\partial t} = \frac{\partial \vP(\vrp,t)}{\partial t} +\nabla\times\vM(\vrp,t)
\ee
\be{10c}
\nabla\cdot\vB(\vrp,t) =0
\ee
\be{10d}
\eps_0\nabla\cdot\vE(\vrp,t) = -\nabla\cdot\vP(\vrp,t)
\ee
\end{subequations}
where the frequency-domain equations have been converted back to the time domain by taking the inverse Fourier transform.  { In other words, the mathematically rigorous solution for all $\vrp$ to the frequency-domain version of (\ref{10}) is given by (\ref{8})--(\ref{9}).  If the surface $S$ of $V$ containing a polarized chunk of continuum material lies in free space outside the chunk, the integrations in (\ref{8}) include the equivalent electric surface charge and current ($-\vnh'\cdot \vP_\w$ and $\vM_\w \times \vnh'$) by means of the surface delta functions in $-\nabla'\cdot\vP_\w$ and $\nabla'\times \vM_\w$; whereas if $S$ lies just inside the chunk of material, these surface contributions are not included.}  To the right-hand sides of (\ref{10b}) and (\ref{10d}) can be added free current $\vJ_f(\vrp,t)$ and free charge $\rho_f(\vrp,t)$, respectively.
\par
With the secondary fields $\vD$ and $\vH$ defined by the constitutive relations
\begin{subequations}
\label{11}
\be{11a}
\vD(\vrp,t) = \eps_0\vE(\vrp,t) + \vP(\vrp,t)
\ee
\be{11b}
\vH(\vrp,t) = \frac{1}{\mu_0}\vB(\vrp,t) - \vM(\vrp,t)
\ee
\end{subequations}
the equations in (\ref{10}) can be rewritten as
\begin{subequations}
\label{12}
\be{12a}
\nabla\times\vE(\vrp,t) +\frac{\partial \vB(\vrp,t)}{\partial t} = 0 
\ee
\be{12b}
\nabla\times\vH(\vrp,t) -\frac{\partial \vD(\vrp,t)}{\partial t} = \vJ_f(\vrp,t)
\ee
\be{12c}
\nabla\cdot\vB(\vrp,t) =0 
\ee
\be{12d}
\nabla\cdot\vD(\vrp,t) = \rho_f(\vrp,t)\,.
\ee
\end{subequations}
\par
The $\vE$ and $\vB$ fields, which have been defined mathematically in the source regions of electric and magnetic polarization $\vP$ and $\vM$ can be related to measurable fields inside a cavity formed by instantaneously removing an infinitesimal volume of $\vP$ and $\vM$ without disturbing the remaining polarization.  For a circular-cylinder infinitesimal volume of length $2b$ and radius $a$, whose axis is aligned with $\vP$, the electric field in the free space at the center of the cylinder differs from the mathematically defined electric field by the amount \cite{Yaghjian-DGF}
\be{13}
\Delta \vE = \frac{\vP}{\eps_0}\left(1-\frac{b}{\sqrt{a^2+b^2}}\right)\,.
\ee
\par
For an infinitesimally narrow cylinder ($a/b\to0$), $\Delta\vE =0$ and the narrow-cylinder (nc) cavity electric field equals $\vE$, the mathematically defined electric field, that is
\be{14}
\vE_c^{\rm nc} =\vE
\ee
where $\vE_c^{\rm nc}$ is measurable (in principle) in the free space of the cavity and provides a method for determining the primary electric field $\vE$.
\par
If $b/a\to0$ so that the cylinder becomes a disk, $\Delta\vE = \vP/\eps_0$ and the thin-disk (td) cavity electric field is given by
\be{15}
\vE_c^{\rm td} =\vE + \frac{\vP}{\eps_0} =\frac{\vD}{\eps_0}\,.
\ee
Thus measuring $\vE_c^{\rm td}$ provides a method for determining the secondary electric field $\vD$ (often called the electric displacement because it is produced by charge separation or ``displacement", as in a capacitor).
\par
Similarly, for an infinitesimal circular cylinder aligned with $\vM$ \cite{Yaghjian-DGF}
\be{16}
\Delta \vB = -\mu_0\vM\frac{b}{\sqrt{a^2+b^2}}
\ee
so that
\be{17}
\vB_c^{\rm nc} = \vB -\mu_0\vM = \mu_0\vH
\ee
\be{1*}
\vB_c^{\rm td} =\vB\,.
\ee
Thus the cavity magnetic fields provide a way to measure the mathematically defined primary magnetic field $\vB$ (often called the magnetic \textit{induction} because the time derivative of its flux through an open surface \textit{induces} an electromotive force around the edge of the surface, as in an inductor) and the related secondary magnetic field $\vH$.
\subsection{\label{MacroscopicContinuum}{Maxwell's equations for macroscopic dipolar continua}}
The ideal dipolar-continuum {Maxwell} equations in (\ref{10}) or (\ref{12}) have been derived assuming that each differential volume element of {$\vP dV$ and $\vM dV$} have the same properties as the continuum as a whole, that is, they are simply infinitesimal chunks of a dipolar continuum with no free space outside the surfaces of the chunks except at the outer surface of the continuum.  However, most natural materials and many metamaterials are comprised of discrete molecules or inclusions separated from one another by a finite distance in free space.  If the operational temporal and spatial bandwidths of the externally applied signals are low enough that these molecules exhibit only electric and magnetic dipole moments with no significant higher order multipole moments, and their average fields vary slowly over electrically small macroscopic volumes $\Delta V$ containing many molecules or inclusions, we will now show that these average fields obey the same {Maxwell} equations as in (\ref{10}) or (\ref{12}).
\par
To simplify the derivation, let $\Delta V$ be a spherical electrically small volume containing a large number of the discrete electric and magnetic dipoles and \textit{assume that the surface $\Delta S$ of $\Delta V$ lies in free space without cutting through any of the dipoles.}\footnote{\lbl{foot5}The polarization densities and fields defined by averaging (at each instant of time $t$) over the macroscopic volumes $\Delta V$ containing discrete numbers of dipoles will not be perfectly continuous functions of position $\vrp$ as individual dipoles are included or not within $\Delta V$ as the center $\vrp$ of $\Delta V$ changes position \cite[pp. 2--3]{Stratton}, \cite{Russakoff}.  These microscopic discontinuities associated with macroscopic-volume averaging over discrete numbers of dipoles can be sufficiently smoothed by various mathematical techniques to allow the curl and divergence of the fields in Maxwell's differential equations to be well-defined.  Alternatively, the volume definitions of curl and divergence \cite{Tai} can be applied directly to the macroscopic volumes.}  (We can imagine displacing slightly a few of the dipoles or part of the spherical surface to realize this assumption; the fractional error in the macroscopic polarization densities and fields introduced by this displacement is on the order of the ratio of the diameter of the dipoles to the diameter of the spherical macroscopic volume $\Delta V$.)  The average electric and magnetic dipole moments of the bound charge and current $\varrho_b$ and $\vj_b$ within $\Delta V$ are given by
\begin{subequations}
\label{21}
\be{21a}
\vP(\vrp,t)  = \frac{1}{\Delta V}\int\limits_{\Delta V} \varrho_b(\vrp+\vrp',t)\vrp' dV'
\ee
\be{21b}
\vM(\vrp,t) = \frac{1}{2\Delta V}\int\limits_{\Delta V} \vrp'\times\vj_b(\vrp+\vrp',t) dV'
\ee
\end{subequations}
where $\vrp$ is the position of the center of the spherical volume $\Delta V$.  We can use the same continuum polarization symbols $\vP$ and $\vM$ defined in (\ref{6'}) (after converting (\ref{6'}) to the time domain by taking the inverse Fourier transform) for the average macroscopic dipole moments in (\ref{21})  because it is assumed that the diameter of $\Delta V$ in (\ref{21}) is much smaller than the smallest significant wavelength in the operational temporal and spatial bandwidths and, thus, the macroscopic volume $\Delta V$ can effectively be used in place of the infinitesimal differential volume element $dV$.  It is assumed that the total bound charge in each $\Delta V$ is zero so that its electric dipole moment is independent of the position of the origin $\vrp$ with respect to $\Delta V$. The magnetic dipole moment of the bound current in $\Delta V$ is dependent upon the position of the origin $\vrp$, unless the electric dipole moment is zero.  However, for $\vrp$ located within $\Delta V$, the dependence of $\vM$ on position is minimal if $\Delta V$ can be made much smaller than the smallest significant wavelength and still contain many discrete dipoles.
\par
It will now be shown that the electric and magnetic fields averaged over the spherical macroscopic volume $\Delta V$ are approximately equal to the average fields within $\Delta V$ of the corresponding ideal continuum.  First, consider the average fields in $\Delta V$ produced by the dipoles contained inside $\Delta V$.  For a finite number $N$ of translationally stationary\footnote{\lbl{foot6}Translating dipoles can be relativistically transformed at each instant of time to a sum of stationary electric and magnetic dipole moments.}  discrete electric and magnetic dipole moments  $\vp_i$ and $\vm_i$ inside the spherical macroscopic volume $\Delta V$ (centered at the position $\vrp$ at time $t$ and whose surface $\Delta S$ lies in free space not intersecting any of the dipoles), the $\vP$ and $\vM$ in (\ref{21}) can also be expressed as
\begin{subequations}
\label{22}
\be{22a}
\vP(\vrp,t)  = \frac{1}{\Delta V}\sum_{i=1}^N \vp_i
\ee
\be{22b}
\vM(\vrp,t) = \frac{1}{\Delta V}\sum_{i=1}^N \vm_i\,.
\ee
\end{subequations}
Let the microscopic electric field of any one of the electric-charge electric dipoles $\vp_i$ be denoted by $\ve_i(\vrp,t)$.  Then the average electric field $\vE^{\rm ins}_{{\rm av},i}$ of this dipole inside the spherical volume $\Delta V$ is defined by
\be{23}
\vE^{\rm ins}_{{\rm av},i} = \frac{1}{\Delta V}\int\limits_{\Delta V}\ve_i(\vrp',t) dV'\,.
\ee
For operational bandwidths that are not so large that the diameter of the macroscopic volume can be a small fraction of a minimum wavelength (and yet contain many dipoles), the electric field $\ve_i$ over the volume $\Delta V$ is dominated by the quasi-electrostatic fields and, thus, the average in (\ref{23}) is simply $\vE^{\rm ins}_{{\rm av},i}=-\vp_i/(3\eps_0)$ \cite[sec. 4.1]{Jackson}.  Summing over the $N$ electric dipoles and using (\ref{22a}) gives
\be{24}
\vE^{\rm ins}_{\rm av} = -\frac{\vP(\vrp,t)}{3\eps_0}
\ee
where $\vE^{\rm ins}_{\rm av}$ is the average over $\Delta V$ of the quasi-electrostatic field of the dipoles inside the $\Delta V$ centered at the position $\vrp$ at time $t$.
Similarly, for the B field averaged over the spherical $\Delta V$ produced by the Amperian magnetic dipoles $\vm_i$ inside $\Delta V$, one finds with the help of (\ref{22b}) that \cite[sec. 5.6]{Jackson}
\be{25}
\vB^{\rm ins}_{\rm av} = \frac{2\mu_0\vM(\vrp,t)}{3}\,.
\ee
{The $2/3$ factor in (\ref{25}) replaces the $-1/3$ factor in (\ref{24}) because the magnetic dipoles are produced by circulating electric currents rather than equal and opposite magnetic charges.}
\par
The average over $\Delta V$ of the quasi-static electric fields produced by the magnetic dipoles inside $\Delta V$ is negligible compared to the average of the quasi-electrostatic fields $\vE^{\rm ins}_{\rm av}$ produced by the electric dipoles inside $\Delta V$.  Likewise, the average over $\Delta V$ of the quasi-static magnetic fields produced by the electric dipoles inside $\Delta V$ is negligible compared to the average of the quasi-magnetostatic fields $\vB^{\rm ins}_{\rm av}$ produced by the magnetic dipoles inside $\Delta V$.
\par
Second, consider the fields averaged over $\Delta V$ produced by the dipoles outside $\Delta V$.  The average over the spherical volume $\Delta V$ of the quasi-electrostatic fields from the electric dipoles outside $\Delta V$ is simply equal to the value of the quasi-electrostatic field at the center of the free-space spherical cavity $\Delta V$; that is \cite[sec. 4.1]{Jackson}
\be{26}
\vE^{\rm out}_{\rm es,av} = \vE^{\rm out}_{\rm es}(\vrp,t)
\ee
where $\vrp$ is at the center of the sphere.  Since the center of the sphere is much further away from the dipoles outside $\Delta V$ than the average distance between the dipoles, the discrete dipoles outside $\Delta V$ can be approximated by an ideal continuum for the sake of determining $\vE^{\rm out}_{\rm es}(\vrp,t)$.  Moreover, this ideal continuum approximation becomes even better for the dipoles so far away from $\Delta V$ that their radiation fields become appreciable to or much larger than the quasi-static fields. Then these distant fields will not vary significantly over the electrically small volume $\Delta V$ and their average fields will also equal the fields at the center of the spherical $\Delta V$.   Consequently, the electric field of the dipoles outside $\Delta V$ averaged over $\Delta V$ is equal to a good approximation to the averaged electric field from the continuum outside $\Delta V$; that is
\be{27}
\vE^{\rm out}_{\rm av} = \vE^{\rm out}
\ee
where $\vE^{\rm out}$ denotes the average electric field in $\Delta V$ produced by the ideal continuum lying outside $\Delta V$.  Similarly,
\be{28}
\vB^{\rm out}_{\rm av} = \vB^{\rm out}\,.
\ee
\par
Adding the inside-dipole average fields in (\ref{24})--(\ref{25}) to the outside-dipole average fields in (\ref{27})--(\ref{28}) for the spherical $\Delta V$, yields the average fields
\begin{subequations}
\label{29}
\be{29a}
\vE_{\rm av} = \vE^{\rm out} - \frac{\vP(\vrp,t)}{3\eps_0}
\ee
\be{29b}
\vB_{\rm av} = \vB^{\rm out}+\frac{2\mu_0\vM(\vrp,t)}{3}
\ee
\end{subequations}
of the discrete dipolar material for bandwidths small enough that  $\Delta V$ can be quasi-static in size and yet contain many dipoles.\footnote{{The $\vP/(3\eps_0)$ term in (\ref{29a}) and (\ref{30a}) also appears in the local-field expression of the Lorentz-Lorenz relation.  However, the local field is the electric field at one electric dipole produced by all the other electric dipoles in random or highly symmetric lattices such as cubic lattices \cite[ch. 16]{Kittel}.  It is not an average of the dipole fields over a volume and for asymmetric lattices the $\vP/(3\eps_0)$ term no longer determines the local field.}} Now, as explained above, $\vE^{\rm out}$ and $\vB^{\rm out}$ are equal to the average electric and magnetic fields in the spherical cavity $\Delta V$ produced by an ideal continuum lying outside $\Delta V$.  These fields are generally called the spherical cavity fields of an ideal continuum and are given in terms of the total fields of the ideal continuum as \cite{Yaghjian-DGF, Yaghjian_AJP-1985}  
\begin{subequations}
\label{30}
\be{30a}
\vE^{\rm out} = \vE + \frac{\vP(\vrp,t)}{3\eps_0}
\ee
\be{30b}
\vB^{\rm out} = \vB -\frac{2\mu_0\vM(\vrp,t)}{3}
\ee
\end{subequations}
which implies from (\ref{29}) that
\begin{subequations}
\label{31}
\be{31a}
\vE_{\rm av}(\vrp,t) = \vE(\vrp,t)
\ee
\be{31b}
\vB_{\rm av}(\vrp,t) = \vB(\vrp,t)\,.
\ee
\end{subequations}
{\em In other words,  the macroscopically averaged fields in discrete dipolar materials or metamaterials 
are approximately equal to the mathematically defined ideal continuum fields if the macroscopically averaged dipolarizations are approximately equal to the continuum dipolarizations and the minimum temporal (free-space) and spatial (medium) wavelengths (call them $\lambda_0^{\rm min}$ and $\lambda^{\rm min}$, respectively) are much larger than the average separation distance ($d$) between the discrete dipoles.  Thus, these macroscopically averaged fields and sources obey the same Maxwell equations in (\ref{10}) or (\ref{12}) as the mathematically defined ideal continuum fields and sources, namely}
\begin{subequations}
\label{12'}
\be{12'a}
\nabla\times\vE(\vrp,t) +\frac{\partial \vB(\vrp,t)}{\partial t} =0
\ee
\be{12'b}
\nabla\times\vH(\vrp,t) -\frac{\partial \vD(\vrp,t)}{\partial t} = \vJ_f(\vrp,t)
\ee
\be{12'c}
\nabla\cdot\vB(\vrp,t) =0 
\ee
\be{12'd}
\nabla\cdot\vD(\vrp,t) = \rho_f(\vrp,t)
\ee
\end{subequations}
where the free macroscopic electric charge and current densities are given in terms of the free microscopic discrete electric charges $q_i$ and current densities as
\begin{subequations}
\label{12''}
\be{12''a}
\rho_f(\vrp,t) = \frac{1}{\Delta V}\sum_{i=1}^N q_i
\ee
\be{12''b}
\vJ_f(\vrp,t)=\frac{1}{\Delta V}\sum_{i=1}^N q_i \vv_i
\ee
\end{subequations}
in which $\vv_i$ is the velocity of the charge $q_i$.  The macroscopic constitutive relations follow from (\ref{11}) as
\begin{subequations}
\label{11'}
\be{11'a}
\vD(\vrp,t) = \eps_0\vE(\vrp,t) + \vP(\vrp,t)
\ee
\be{11'b}
\vH(\vrp,t) = \frac{1}{\mu_0}\vB(\vrp,t) - \vM(\vrp,t)\,.
\ee
\end{subequations}
\par
If we denote the maximum temporal and spatial wave numbers by $k_0^{\rm max}=2\pi/\lambda_0^{\rm min}$ and $k^{\rm max}=2\pi/\lambda^{\rm min}$, respectively, then the discrete dipolar medium can be considered a macroscopic continuum that approximates an ideal continuum if both $k_0^{\rm max}d\ll 1$ and  $k^{\rm max}d\ll 1$.  This result, which is one of the cornerstones of classical electromagnetics, has also been confirmed in recent articles that rigorously analyze three-dimensional cubic arrays of inclusions, which are separated from one another in free space, taking into account both temporal and spatial dispersion \cite{YAS1, YAS2, Yaghjian-RS_boundaryconditions}.   It is emphasized that this result is derived and, in general, holds only if the surfaces of the infinitesimal volumes used to define the dipolarization densities in ideal continua and the surfaces of the finite-size macroscopic volumes used to define the dipolarization densities in discrete-dipole macroscopic continua lie in free space so as not to intersect any of the charge and current that produces the electric and magnetic dipoles.
\par
I have not been able to find elsewhere the relatively simple, yet rigorous proof given here that averaging the mi\-cro\-sco\-pic fields of discrete dipoles over macroscopic volumes leads to the same Maxwell equations and cavity fields obtained for mathematically defined fields in ideal continuous dipolar media.  For example, the widely referenced ``theory of electrons" used by Lorentz \cite{Lorentz} and Rosenfeld \cite{Rosenfeld} to obtain the macroscopic Maxwell equations begins by taking the spatial average of the microscopic equations in (\ref{1}) over fixed macroscopic volumes ${\cal V}$ that cut through the charges and dipoles to get
\begin{subequations}
\label{LR1}
\be{LR1a}
\nabla\times{\vcE}(\vrp,t) +\frac{\partial {\vcB}(\vrp,t)}{\partial t} =0
\ee
\be{LR1b}
\frac{1}{\mu_0}\nabla\times{\vcB}(\vrp,t) -\eps_0\frac{\partial {\vcE}(\vrp,t)}{\partial t} = {\vcJ}(\vrp,t)
\ee
\be{LR1c}
\nabla\cdot{\vcB}(\vrp,t) =0
\ee
\be{LR1d}
\eps_0\nabla\cdot{\vcE}(\vrp,t) = {\lrho}(\vrp,t)
\ee
\end{subequations}
where the macroscopic average of $\ve(\vrp,t)$, for example, is defined as
\be{LR2}
{\vcE}(\vrp,t) =\frac{1}{\cal V}\int\limits_{\cal V} \ve(\vrp+\vrp',t) d{\cal V}'
\ee
and similarly for the other fields and source densities.
The macroscopic equations in (\ref{LR1}) hold exactly for any size volume ${\cal V}$ (not just electrically small quasi-static volumes) as long as the limits of integration of the volume ${\cal V}$ {in (\ref{LR2})} stay fixed for each $\vrp$ and thus the surface ${\cal S}$ of ${\cal V}$, unlike the surface $\Delta S$ of the electrically small quasi-static macroscopic volumes $\Delta V$ used above,  will cut through the charges and dipoles as $\vrp$ is varied.  Then, for sufficiently small ${\cal V}$, both Lorentz and Rosenfeld claim that the average ``bound" current and charge on the right-hand sides of (\ref{LR1b}) and (\ref{LR1d}) can be approximated by
\be{LR3}
{\vcJ}(\vrp,t) \approx \frac{\partial {\vcP}(\vrp,t)}{\partial t} +\nabla\times{\vcM}(\vrp,t)
\ee
and
\be{LR4}
{\lrho}(\vrp,t)\approx -\nabla\cdot{\vcP}(\vrp,t)
\ee
where
\begin{subequations}
\label{LR5}
\be{LR5a}
{\vcP}(\vrp,t)\approx\frac{1}{\cal V}\int\limits_{\cal V}\varrho_b(\vrp+\vrp',t)\vrp' d{\cal V}' \approx \frac{1}{\cal V}\sum_{i=1}^N \vp_i  
\ee
\be{LR5b}
{\vcM}(\vrp,t) \approx \frac{1}{2 {\cal V}}\int\limits_{\cal V} \vrp'\times\vj_b(\vrp+\vrp',t) d{\cal V}'\approx \frac{1}{\cal V}\sum_{i=1}^N \vm_i
\ee
\end{subequations}
with $\vp_i$ and $\vm_i$ equal to the electric and magnetic dipole moments of the molecules within ${\cal V}$.  However, if one inserts the integral definitions of ${\vcP}(\vrp,t)$ and ${\vcM}(\vrp,t)$ from (\ref{LR5}) into (\ref{LR3}) and then the resulting ${\vcJ}(\vrp,t)$ into (\ref{LR1b}), one finds with the use of vector-dyadic identities that ${\vcJ}(\vrp,t)=(1/{\cal V})\int_{\cal V} \vj_b(\vrp+\vrp',t) d{\cal V}'$ holds in general only if the surface of ${\cal V}$ lies in free space and does not cut through $\vj_b(\vrp+\vrp',t)$; that is, the ${\cal V}$ in (\ref{LR5}) cannot be the same as the ${\cal V}$ in (\ref{LR2}) and thus the validity of equations (\ref{LR1})--(\ref{LR5}) is uncertain.  This is dramatically illustrated by choosing $\varrho_b = -\nabla\cdot\vcP$ and $\vj_b = \partial\vcP/\partial t +\nabla\times\vcM$ in (\ref{LR5}).
\par
If one ignores this uncertainty and chooses the same ${\cal V}$ in (\ref{LR5}) as in (\ref{LR1})--(\ref{LR2}), the Lorentz-Rosenfeld macroscopic equations in (\ref{LR1})--(\ref{LR5}) (which are also obtained by Van Vleck \cite[sec. 3]{Vleck}) can lead to large variations in the average macroscopic polarizations and fields over microscopic distances (not just the microscopic variations, mentioned in Footnote 6, associated with $\Delta V$).  Consider, for example, the surface ${\cal S}$ cutting through discrete electric dipoles composed of extremely large equal and opposite charges separated by extremely small distances.  Then both the electric polarization and electric field averaged over the macroscopic volume ${\cal V}$ can have extremely large variations over distances comparable to the charge separation of the dipoles, and these variations approach infinite values for point dipoles with nonzero dipole moments.  Moreover, with ${\cal S}$ cutting through the dipoles, the average of the microscopic fields within a cavity ${\cal V}$ will not generally be a good approximation to the average cavity fields of the ideal dipolar continuum equations.
\par  
These large unphysical fluctuations in the Lorentz-Rosenfeld macroscopic equations for dipolar continua can be reduced by using smooth test functions or ensemble averaging instead of fixed macroscopic volumes ${\cal V}$ that cut through the dipoles {\cite{Russakoff}, \cite[chs. 5--6]{Robinson}, \cite[sec. 6.6]{Jackson}}.   {\em Nevertheless, once the smooth macroscopic Maxwell equations are found, it is still necessary for many theoretical, numerical, and experimental purposes (such as deriving the macroscopic power and energy relations from the microscopic equations) to determine the conditions under which the cavity fields of these smooth ideal continuum equations are to a good approximation equal to the average of the microscopic cavity fields of the discrete dipoles.  This determination entails introducing macroscopic volumes $\Delta V$, whose surfaces $\Delta S$ do not intersect the microscopic dipoles, and performing a proof similar to the one given above in the main body of this section.}  The necessity of using macroscopic averaging volumes with surfaces that do not intersect the sources to obtain physically and mathematically robust, unambiguous macroscopic {Maxwell} continuum equations is further confirmed by the rigorous analysis of spatially dispersive metamaterial arrays \cite{YAS1, YAS2}.  De Groot and Suttorp, in their book on the foundations of electrodynamics \cite{Suttorp}, average over discrete {\textit{undivided}} ``stable groups" of point charges that contain electric and magnetic dipoles (and higher order multipoles).  Also, Landau and Lifshitz, in defining macroscopic polarization \cite[secs. 6 and 29]{LLP}, state that the surfaces of the macroscopic volumes must enclose the dipoles {\textit{but nowhere enter them}}.
\section{\label{PTDC}{POYNTING'S THEOREM FOR DIPOLAR CONTINUA}}
Having determined Maxwell's equations for both an ideal and a macroscopic continuum, we shall derive Poynting's theorem in a way that reveals that the integral of the Poynting vector over a closed surface equals the instantaneous power flow in these two types of continua.
\subsection{\label{PTIC}{Poynting's theorem for ideal dipolar continua}}
It was proven in Section \ref{PTcc} that the total electromagnetic power at time $t$ entering a closed surface $S$ lying in free space is given by the integral over $S$ of the microscopic Poynting vector; specifically from (\ref{5'}) 
\be{32}
P(t) = -\int\limits_S \vnh\cdot[\ve(\vrp,t)\times\vh(\vrp,t)] dS 
\ee
where $\vnh$ is the unit normal pointing away from the volume $V$ enclosed by $S$, and $\vh$ has been substituted for $\vb/\mu_0$ because $S$ is in free space.  We showed in Section \ref{IdealContinuum} that the fields in an ideal dipolar continuum satisfied the Maxwell equations in (\ref{12}).  From these equations it follows that the components of $\vE$ and $\vH$ tangential to an interface between free space and the polarized material are continuous, provided there are no equivalent polarization surface currents at the interface of the polarized material, that is, no delta functions in $\vP$ or $\vM$ at the interface.  {Such delta functions can occur for material with constitutive parameters that have zero or infinite values \cite{Yaghjian_ExtremeBC}.  Even if we ignore these extreme values as being unrealizable, finite ($\neq 0$ or $\infty$) constitutive parameters that are strongly spatially dispersive can also exhibit delta functions in $\vP$ and $\vM$ at the interface \cite{YAS1, YAS2}.  (The {usual} Poynting vector in strongly spatially dispersive material does not generally represent energy flow \cite[p. 361]{LLP}, \cite{CSA}.)}   If we  assume low enough frequencies that spatial dispersion is negligible, then the tangential components of $\vE$ and $\vH$ are continuous across the free-space/dipolar-material interface.  This implies that if we remove an infinitesimally thin shell of material containing $S$ (everywhere that the closed surface $S$ passes through a dipolar continuum), the value of $\vnh\cdot(\vE\times\vH)$ will be continuous across the free-space shell.  Moreover, since the continuum fields equal the microscopic fields in the resulting free-space shell, that is
\be{33}
\vnh\cdot(\vE\times\vH) = \vnh\cdot(\ve\times\vh)
\ee
in the free-space shell, one finds from (\ref{32}) that the total instantaneous power flowing across a closed surface $S$ in an ideal dipolar continuum is given by the integration of the continuum Poynting vector $P(t) = -\int\limits_S \vnh\cdot[\vE(\vrp,t)\times\vH(\vrp,t)] dS$, or with the aid of equations (\ref{12a}) and (\ref{12b})
\be{34}
P(t) = -\int\limits_S \vnh\cdot[\vE(\vrp,t)\times\vH(\vrp,t)] dS =\int\limits_V \left[\frac{\partial\vD(\vrp,t)}{\partial t}\cdot\vE(\vrp,t)
+ \frac{\partial\vB(\vrp,t)}{\partial t}\cdot\vH(\vrp,t) \right] dV\,.
\ee
Thus, we have determined that in an ideal continuum the integral of the continuum Poynting vector over a closed surface $S$ of a volume $V$ exactly equals the instantaneous power flow across that surface, provided the spatial dispersion is not strong enough over the operational bandwidths to produce {delta functions in $\vP$ and $\vM$} at a hypothetical interface $S$ between the polarized material and free space.  Furthermore, the surface integral of the continuum Poynting vector equals the corresponding volume integral of the fields on the right-hand side of (\ref{34}).  If the material is linear and characterized by a frequency independent permittivity $\eps(\vrp)$ and permeability $\mu(\vrp)$ over the operational bandwidth, then (\ref{34}) becomes
\be{35}
P(t) = -\int\limits_S \vnh\cdot[\vE(\vrp,t)\times\vH(\vrp,t)] dS =\frac{1}{2}\frac{\partial}{\partial t}\int\limits_V \left[\eps(\vrp)|\vE(\vrp,t)|^2 + \mu(\vrp)|\vH(\vrp,t)|^2 \right] dV
\ee
which implies that in a nondispersive, scalar, linear magnetodielectric material, $[\eps(\vrp)|\vE(\vrp,t)|^2 +\mu(\vrp)|\vH(\vrp,t)|^2]/2$ is the electromagnetic energy density stored in the $\vE$ and $\vH$ fields.
\par
It should be noted that in some contexts bianisotropic media are considered to be spatially dispersive because the electric and magnetic fields in bianisotropic media are proportional to the curls of the magnetic and electric fields, respectively.  However, since this weak spatial dispersion maintains continuous $\vnh\cdot(\vE\times\vH)$ across interfaces between the bianisotropic continua and free space, for the purposes of this paper, we can include the weak spatial dispersion of bianisotropic media within the context of spatially nondispersive media; {see Footnote 16}.
\subsection{\label{PTMC}{Poynting's theorem for macroscopic dipolar continua}}
The argument leading to Poynting's theorem in an ideal dipolar continuum and its inter\-pretation in terms of power flow can be applied with a couple of modifications to a macroscopic dipolar continuum comprised of discrete electric and magnetic dipoles.  The first modification is that the volume $V$ with surface $S$ to which Poynting's theorem is applied should be no smaller than a macroscopic volume $\Delta V$ which is electrically small but still contains so many dipoles that the averaging of the fields over $\Delta V$ gives to a good approximation the continuum fields satisfying Maxwell's equations in (\ref{10}) or (\ref{12}) and (\ref{12'}); see Section \ref{MacroscopicContinuum}.  
\par
The second modification involves a transition layer {\cite[p. 271]{Drude}, \cite[sec. 1.13]{Stratton}} of finite thickness at an interface between free space and the macroscopic dipolar continuum \cite{YAS1, YAS2, Yaghjian-RS_boundaryconditions}.   Unlike an ideal dipolar continuum, all the components of the $\vE$ and $\vH$ fields in a macroscopic dipolar continuum without the interface are not generally equal to the fields close to an interface between free space and the macroscopic continuum.  However, the tangential $\vE$ and $\vH$ fields with and without the interface are nearly continuous across a transition layer of thickness $\delta$ containing the interface.  Fortunately, under the conditions that the discrete dipolar material behaves as a continuum, namely  $k_0^{\rm max}d\ll 1$ and  $k^{\rm max}d\ll 1$ (see Section \ref{MacroscopicContinuum}), analytical and numerical results with discrete dipolar arrays indicate that the thickness $\delta$ of the transition layer is on the order of {a few average separation distances} $d$ of the dipoles \cite{S&T, S&K}{.}\footnote{{Computations in front of planar arrays of point dipole sources separated by a distance $d\ll\lambda_0^{\rm min}$ show that the fractional variations in the fields caused by the localization of the point dipoles are no greater than about $1\%$ for a distance $z=d$ in front of the plane of the array; and the fractional variations rapidly decrease with $z>d$.  In other words, $\vnh\cdot(\ve\times\vh)$ at the center of a free-space shell just two lattice layers thick ($\delta =2d$) is equal to $\vnh\cdot(\vE\times\vH)$ to an accuracy better than a few percent with the accuracy rapidly improving for $\delta > 2d$ (and the integrations over $S$ of these two Poynting vectors  agree to appreciably greater accuracy).  Also, extensive calculations with three-dimensional {dipole} arrays show that they are well approximated by continua as the lattice spacing $d$ becomes less than about a tenth of a wavelength \cite{S&Y}.}}  This means that the infinitesimal thickness of the free-space shell about $S$ that was chosen to derive the results for the Poynting vectors in Section \ref{PTIC} for the ideal continuum can be made equal to the thickness $\delta$ of the transition layer without appreciably changing the average fields within the volume $V$.  Consequently, the equation (\ref{33}) holds to a good approximation for $S$ at the center of the $\delta$-thick free-space shell that removes a discrete number of dipoles {(see Fig. \ref{fig2})}, and the total instantaneous power flowing across the closed surface $S$ in free space just outside the volume $V$ containing a discrete number of dipoles is given to a good approximation by (\ref{34}), that is 
\be{36}
P(t) \approx  -\int\limits_S \vnh\cdot[\vE(\vrp,t)\times\vH(\vrp,t)] dS =\int\limits_V \left[\frac{\partial\vD(\vrp,t)}{\partial t}\cdot\vE(\vrp,t)
+ \frac{\partial\vB(\vrp,t)}{\partial t}\cdot\vH(\vrp,t) \right] dV\,.
\ee
The better the discrete distribution of dipoles approximates a continuum, the thinner the transition layer becomes ($\delta\to0$ for an ideal continuum) and the more accurate is the approximation in (\ref{36}) for the macroscopic continuum.
\begin{figure}[ht]
\begin{center}
\includegraphics[width =4.0in]{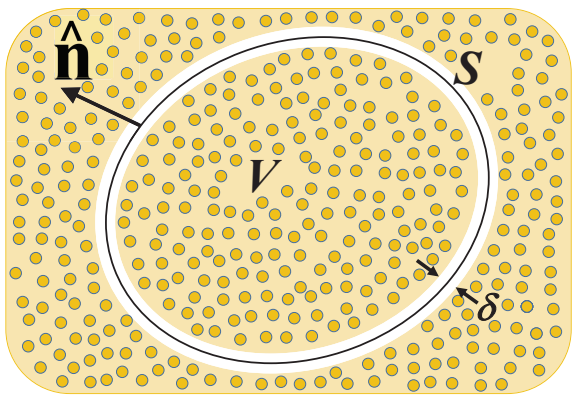}
\end{center}
\caption{\label{fig2}Volume $V$ (or $\Delta V$) with its surface $S$ {(or $\Delta S$)} centered in a free-space shell of thickness $\delta$ enclosing a large number of discrete electric and magnetic dipoles.}
\end{figure}
%
\par
The most important aspect of (\ref{34}) for the ideal continuum and in (\ref{36}) for the macroscopic continuum is that both the surface and volume integrals in (\ref{34}) and (\ref{36}) are equal to the instantaneous power flow $P(t)$ through the closed surface $S$ into the volume $V$, exactly for the ideal continuum and approximately for the macroscopic continuum, respectively.  Although free-space shells are invoked to prove (\ref{34}) and (\ref{36}), the fields in (\ref{34}) and (\ref{36}) are the continuum fields and thus (\ref{34}) holds exactly within the ideal continuum,  and (\ref{36}) holds approximately in the macroscopic continuum since $\vnh\cdot(\vE\times\vH)$ is continuous across a free-space/dipolar-material interface under the conditions specified above and satisfied by most dipolar metamaterial continua as well as natural dipolar material continua.
\section{\label{ER}{ENERGY RELATIONS FOR MACROSCOPIC DIPOLAR CONTINUA}}
If the volume $V$ in (\ref{36}) is chosen to be a macroscopic volume $\Delta V$ that is electrically small but large enough to contain a great number of discrete dipoles, then (\ref{36}) becomes
\be{36'}
\Delta P(t) \approx  -\int\limits_{\Delta S} \vnh\cdot[\vE(\vrp,t)\times\vH(\vrp,t)] dS =\int\limits_{\Delta V} \left[\frac{\partial\vD(\vrp,t)}{\partial t}\cdot\vE(\vrp,t)
+ \frac{\partial\vB(\vrp,t)}{\partial t}\cdot\vH(\vrp,t) \right] dV\,.
\ee
We can also express the power $\Delta P(t)$ in terms of the microscopic fields and sources within $\Delta V$.  This can be done by reinstating the surface $\Delta S$ of $\Delta V$ in the middle of a free-space shell of transition layer thickness $\delta$ that removes a discrete number of dipoles, as explained in Section \ref{PTMC}, where the volume $\Delta V$ contains a large number of discrete electric and magnetic dipoles.  Since $\Delta S$ is now in free space a distance $\delta/2$ away from the nearest dipole, the average macroscopic $\vE$ and $\vH$ fields there are approximately equal to the microscopic $\ve$ and $\vh =\vb/\mu_0$ fields, as explained in Section \ref{PTDC}.  Thus, from (\ref{36'}) and (\ref{5'}),  the power $\Delta P(t)$ can be written as
\bea{39}
\Delta P(t) &\approx& -\int\limits_{\Delta S} \vnh\cdot[\ve(\vrp,t)\times\vb(\vrp,t)/\mu_0] dS \nonumber\\ 
&=&   \int\limits_{\Delta V} \vj(\vrp,t)\cdot\ve(\vrp,t) dV  + \frac{1}{2}\frac{d}{dt}\int\limits_{\Delta V}\left[\eps_0|\ve(\vrp,t)|^2 +\frac{1}{\mu_0}|\vb(\vrp,t)|^2\right] dV\nonumber\\
&\approx& \int\limits_{\Delta V} \left[\frac{\partial\vD(\vrp,t)}{\partial t}\cdot\vE(\vrp,t)
+ \frac{\partial\vB(\vrp,t)}{\partial t}\cdot\vH(\vrp,t) \right] dV\nonumber\\
&=& \int\limits_{\Delta V} \left[\frac{\partial\vP(\vrp,t)}{\partial t}\cdot\vE(\vrp,t)
- \frac{\partial\vB(\vrp,t)}{\partial t}\cdot\vM(\vrp,t) \right] dV\nonumber\\ 
&& \hspace{30mm}+ \, \frac{1}{2}\frac{d}{dt}\int\limits_{\Delta V}\left[\eps_0|\vE(\vrp,t)|^2 +\frac{1}{\mu_0}|\vB(\vrp,t)|^2\right] dV\,.
\eea
This equation reveals that
\bea{39'}
\int\limits_{\Delta V} \left[\frac{\partial\vP(\vrp,t)}{\partial t}\cdot\vE(\vrp,t)
- \frac{\partial\vB(\vrp,t)}{\partial t}\cdot\vM(\vrp,t) \right] dV \hspace{50mm}\nonumber\\
\approx \int\limits_{\Delta V} \vj(\vrp,t)\cdot\ve(\vrp,t) dV 
+ \frac{1}{2}\frac{d}{dt}\int\limits_{\Delta V}\left\{\eps_0\left[|\ve(\vrp,t)|^2 -{|}\vE(\vrp,t)|^2\right] +\frac{1}{\mu_0}\left[|\vb(\vrp,t)|^2 - |\vB(\vrp,t)|^2\right]\right\} dV\,.
\eea
The last integral in (\ref{39'}) can be simplified using the identities
\begin{subequations}
\label{40}
\be{40a}
\int\limits_{\Delta V}|\ve -\vE|^2 dV = \int\limits_{\Delta V}(|\ve|^2 +|\vE|^2 -2\ve\cdot\vE)dV =\int\limits_{\Delta V}(|\ve|^2 -|\vE|^2) dV
\ee
\be{40b}
\int\limits_{\Delta V}|\vb -\vB|^2 dV = \int\limits_{\Delta V}(|\vb|^2 +|\vB|^2 -2\vb\cdot\vB)dV =\int\limits_{\Delta V}(|\vb|^2 -|\vB|^2) dV
\ee
\end{subequations}
which hold because $\vE$ and $\vB$ within $\Delta V$ are practically uniform over $\Delta V$ and these primary macroscopic fields are defined simply in terms of the microscopic fields as
\begin{subequations}
\label{41}
\be{41a}
\vE(\vrp\in\Delta V,t) = \frac{1}{\Delta V}\int\limits_{\Delta V}\ve(\vrp,t) dV 
\ee
\be{41b}
\vB(\vrp\in\Delta V,t) = \frac{1}{\Delta V}\int\limits_{\Delta V}\vb(\vrp,t) dV\,.
\ee
\end{subequations}
Moreover, the microscopic fields within $\Delta V$ from the dipoles outside of $\Delta V$ are practically equal to the macroscopic fields from the dipoles outside of $\Delta V$.  That is, both the microscopic and macroscopic fields within $\Delta V$ produced by the dipoles outside $\Delta V$ are approximately equal to the cavity fields in a cavity $\Delta V$ of the corresponding ideal continuum and thus $(\ve^{\rm out}-\vE^{\rm out}) \approx 0$ and $(\vb^{\rm out}-\vB^{\rm out}) \approx 0$, where the superscripts ``out" denote fields within $\Delta V$ from dipoles \underline{out}side $\Delta V$.  Therefore,
\begin{subequations}
\label{42}
\be{42a}
|\ve -\vE|^2 \approx |\ve^{\rm ins} -\vE^{\rm ins}|^2
\ee
\be{42b}
|\vb -\vB|^2 \approx |\vb^{\rm ins} -\vB^{\rm ins}|^2
\ee
\end{subequations}
where the superscripts ``ins" denote the fields within $\Delta V$ produced by the dipoles \underline{ins}ide $\Delta V$.  By definition
\begin{subequations}
\label{41ins}
\be{41insa}
\vE^{\rm ins}(\vrp\in\Delta V,t) = \frac{1}{\Delta V}\int\limits_{\Delta V}\ve^{\rm ins}(\vrp,t) dV 
\ee
\be{41insb}
\vB^{\rm ins}(\vrp\in\Delta V,t) = \frac{1}{\Delta V}\int\limits_{\Delta V}\vb^{\rm ins}(\vrp,t) dV\,.
\ee
\end{subequations} 
Combining (\ref{42}) with (\ref{40}), inserting the result into (\ref{39'}), then performing the volume integration on the left-hand side of (\ref{39'}) by using the fact that the macroscopic sources and fields are nearly uniform over the electrically small macroscopic volume $\Delta V$, one obtains
\be{43}
 \frac{\partial\vP}{\partial t}\cdot\vE
- \frac{\partial\vB}{\partial t}\cdot\vM  
\approx \frac{1}{\Delta V}\int\limits_{\Delta V} \vj\cdot\ve dV 
+ \frac{1}{2\Delta V}\frac{d}{dt}\int\limits_{\Delta V}\left[\eps_0|\ve^{\rm ins} -\vE^{\rm ins}|^2 +\frac{1}{\mu_0}|\vb^{\rm ins} - \vB^{\rm ins}|^2\right] dV\,.
\ee
Integration of this last equation from time $t_0$ to time $t$ gives
\bea{44}
&&\hspace{-8mm}\int\limits_{t_0}^t \left[\frac{\partial\vP(\vrp,t')}{\partial t'}\cdot\vE(\vrp,t')
- \frac{\partial\vB(\vrp,t')}{\partial t'}\cdot\vM(\vrp,t') \right] dt'\nonumber\\
&&\approx \frac{1}{\Delta V} \int\limits_{t_0}^t\int\limits_{\Delta V} \vj\cdot\ve dVdt' 
+ \frac{1}{2\Delta V}\int\limits_{\Delta V}\left[\eps_0|\ve^{\rm ins} -\vE^{\rm ins}|^2 +\frac{1}{\mu_0}|\vb^{\rm ins} - \vB^{\rm ins}|^2\right]_{t_0}^t dV\nonumber\\
&&\hspace{-8mm}=  \frac{1}{\Delta V} \int\limits_{t_0}^t\int\limits_{\Delta V} \vj\cdot\ve dVdt'\\ 
&&\hspace{-7mm}+ \frac{1}{2\Delta V}\int\limits_{\Delta V}\left[\eps_0|\ve^{\rm ins}(\vrp,t) -\vE^{\rm ins}(\vrp,t)|^2 +\frac{1}{\mu_0}|\vb^{\rm ins}(\vrp,t) - \vB^{\rm ins}(\vrp,t)|^2 -\eps_0|\ve^{\rm ins}(\vrp,t_0)|^2 -\frac{1}{\mu_0}|\vb^{\rm ins}(\vrp,t_0)|^2\right] dV\nonumber 
\eea
because $\vE^{\rm ins}(\vrp,t_0)=0$ and $\vB^{\rm ins}(\vrp,t_0)=0$.
\par
This result shows that the change in macroscopic polarization energy in $\Delta V$ determined by the left-hand side of (\ref{44}) is not just equal to the work done by the microscopic electric field on the microscopic current density (the mechanical energy represented by the double integral on the right-hand side of (\ref{44})) because of the additional change in field energy on the right-hand sides of (\ref{44}).  { We will show next that this additional field energy produced by the dipoles inside  any one macroscopic volume $\Delta V$ is confined to that $\Delta V$ and thus behaves as an isolated capacitive and inductive energy of the system of dipoles inside $\Delta V$.} 
\par
Depending on the properties of the discrete dipoles, the last volume integral in (\ref{44}) may or may not be greater than or equal to zero. To better understand this, it helps to give a physical interpretation of the integral of the ``inside" fields in (\ref{44}).  The fields $(\ve^{\rm ins}, \vb^{\rm ins})$ are the microscopic fields produced by the discrete dipoles located within the electrically small macroscopic volume $\Delta V$, whereas $(\vE^{\rm ins}, \vB^{\rm ins})$ are the fields produced by the corresponding continuum polarizations $(\vP,\vM)$ replacing the discrete dipole moments within $\Delta V$.  (For a spherical $\Delta V$, $\vE^{\rm ins} \approx -\vP/(3\eps_0)$ and $\vB^{\rm ins} \approx 2\mu_0\vM/3$.)   Therefore the  fields $\ve^{\rm ins}$ and $\vE^{\rm ins}$ both satisfy the quasi-electrostatic Maxwell equations with discrete electric dipole moments and continuous electric polarization in $\Delta V$, respectively. Likewise, the  fields $\vb^{\rm ins}$ and $\vB^{\rm ins}$ both satisfy the quasi-magnetostatic Maxwell equations with discrete magnetic dipole moments and continuous magnetic polarization (magnetization) in $\Delta V$, respectively.  For observation points outside $\Delta V$, the two pairs of fields are approximately equal, that is,  $\ve^{\rm ins}(\vrp,t)\approx \vE^{\rm ins}(\vrp,t)$ and $\vb^{\rm ins}(\vrp,t)\approx \vB^{\rm ins}(\vrp,t)$ for $\vrp\notin\Delta V$, whereas for observation points within $\Delta V$, the two pairs of fields can differ greatly.  Consequently
\bea{45}
\frac{1}{2}\int\limits_{V_\infty}\left[\eps_0|\ve^{\rm ins}(\vrp,t) -\vE^{\rm ins}(\vrp,t)|^2 +\frac{1}{\mu_0}|\vb^{\rm ins}(\vrp,t) - \vB^{\rm ins}(\vrp,t)|^2\right] dV\nonumber\\ \approx \frac{1}{2}\int\limits_{\Delta V}\left[\eps_0|\ve^{\rm ins}(\vrp,t) -\vE^{\rm ins}(\vrp,t)|^2 +\frac{1}{\mu_0}|\vb^{\rm ins}(\vrp,t) - \vB^{\rm ins}(\vrp,t)|^2\right] dV
\eea
where $V_\infty$ denotes all space.  Thus, the electric-field integral over $\Delta V$ can be viewed as local capacitive energy of the electric dipoles within $\Delta V$ and the magnetic-field integral over $\Delta V$ can be viewed as local inductive energy of the magnetic dipoles within $\Delta V$, at every instant of time $t$:
\begin{subequations}
\lbl{WCL}
\be{WC}
\Delta W_C(\vrp,t) =\frac{1}{2}\int\limits_{\Delta V}\eps_0\left[|\ve^{\rm ins}(\vrp,t) -\vE^{\rm ins}(\vrp,t)|^2 \right] dV
\ee
\be{WL}
\Delta W_L(\vrp,t) =\frac{1}{2}\int\limits_{\Delta V}\frac{1}{\mu_0}\left[|\vb^{\rm ins}(\vrp,t) -\vB^{\rm ins}(\vrp,t)|^2 \right] dV
\ee
\end{subequations}
where the $\vrp$ in $\Delta W_C(\vrp,t)$ and $\Delta W_L(\vrp,t)$ indicates the position of the macroscopic volume $\Delta V$.
The subtraction of $\vE^{\rm ins}$ and $\vB^{\rm ins}$ from $\ve^{\rm ins}$ and $\vb^{\rm ins}$, respectively, in (\ref{45}) removes the fields contributed by the truncation of the volume $\Delta V$ at the inner surface of the free-space shell shown in Fig. \ref{fig2}.  {These difference fields produced by the dipole moments inside $\Delta V$ do not extend across $\Delta S$.}  The energy in these inside fields plus the mechanical energy supplied by the microscopic electric field $\ve$ acting on the microscopic current $\vj$ in $\Delta V$ through the $\vj\cdot\ve$ integral on the right-hand side of (\ref{44}) (see Section \ref{Energy}) is equal to the change in macroscopic polarization energy density in $\Delta V$ determined by the first integral in (\ref{44}).
\par
If we divide the electric current $\vj$ into the electric current $\vj_e$ of the electric dipoles and the electric current $\vj_b$ of the magnetic dipoles, such that $\vj =\vj_e +\vj_b$, and define electric and magnetic polarization energy densities as
\begin{subequations}
\lbl{ped}
\be{peda}
U_e(\vrp,t) =  \frac{1}{\Delta V} \int\limits_{t_0}^t\int\limits_{\Delta V} \vj_e\cdot\ve dVdt' 
+ \frac{1}{2\Delta V}\int\limits_{\Delta V}\eps_0|\ve^{\rm ins}(\vrp,t) -\vE^{\rm ins}(\vrp,t)|^2 dV 
= \frac{1}{\Delta V}\left[\Delta W_{j_ee}(\vrp,t) +\Delta W_C(\vrp,t)\right]
\ee
\be{pedb}
U_b(\vrp,t) =  \frac{1}{\Delta V} \int\limits_{t_0}^t\int\limits_{\Delta V} \vj_b\cdot\ve dVdt' 
+ \frac{1}{2\Delta V}\int\limits_{\Delta V}\frac{1}{\mu_0}|\vb^{\rm ins}(\vrp,t) - \vB^{\rm ins}(\vrp,t)|^2  dV
= \frac{1}{\Delta V}\left[\Delta W_{j_be}(\vrp,t) +\Delta W_L(\vrp,t)\right]
\ee
\end{subequations}
where
\begin{subequations}
\lbl{5WCL'}
\be{5WCL'a}
\Delta W_{je}(\vrp,t) = \Delta W_{j_ee}(\vrp,t)+\Delta W_{j_be}(\vrp,t) = \int\limits_{t_0}^t \int\limits_{\Delta V} \vj(\vrp,t')\cdot\ve(\vrp,t') dV dt'
\ee
with
\be{5WCL'b}
\Delta W_{j_ee}(\vrp,t) = \int\limits_{t_0}^t \int\limits_{\Delta V} \vj_e(\vrp,t')\cdot\ve(\vrp,t') dV dt'
\ee
\be{5WCL'c}
\Delta W_{j_be}(\vrp,t) = \int\limits_{t_0}^t \int\limits_{\Delta V} \vj_b(\vrp,t')\cdot\ve(\vrp,t') dV dt'
\ee
\end{subequations}
{then the change in macroscopic polarization energy density in (\ref{44}) can be rewritten as}   
\be{44''}
\int\limits_{t_0}^t \left[\frac{\partial\vP(\vrp,t')}{\partial t'}\cdot\vE(\vrp,t')
- \frac{\partial\vB(\vrp,t')}{\partial t'}\cdot\vM(\vrp,t') \right] dt' 
\approx  U_e(\vrp,t) -U_e(\vrp,t_0)+ U_b(\vrp,t) -U_b(\vrp,t_0)\,.
\ee
If the initial energies of formation\footnote{\lbl{formation}{The energy of formation of each dipole is the work needed to differentially and quasi-statically assemble the charges or currents in free space from an infinite separation distance.  It can be rewritten in terms of the integration in free space over the static electric and magnetic field of each electric or magnetic dipole, respectively.  The energy of formation of a permanent electric dipole does not change if the separation distance of its charges do not change.  The energy of  a ``permanent" magnetic dipole does not change if its current does not change.}} of the electric and magnetic dipoles do not change or they get greater with time (that is, no energy can be extracted from the initial energy of the dipoles), then in passive material
\be{subv6}
 U_e(\vrp,t) -U_e(\vrp,t_0)+ U_b(\vrp,t) -U_b(\vrp,t_0)\ge 0
\ee
and we have 
\be{44up}
\int\limits_{t_0}^t \left[\frac{\partial\vP(\vrp,t')}{\partial t'}\cdot\vE(\vrp,t')
- \frac{\partial\vB(\vrp,t')}{\partial t'}\cdot\vM(\vrp,t') \right] dt' 
\approx  U_e(\vrp,t) -U_e(\vrp,t_0)+ U_b(\vrp,t) -U_b(\vrp,t_0)\ge0.
\ee
{To clearly demonstrate this inequality, chose a spherical volume $\Delta V$ and enclose the dipoles in $\Delta V$ with a perfect conductor so that the electric and magnetic fields in $\Delta  V$ will equal $\ve^{\rm ins}-\vE^{\rm ins}$ and $\vb^{\rm ins}-\vB^{\rm ins}$, respectively.  Then an external source exerting nonelectromagnetic forces equal to $\ve(\vrp,t)$ on the dipole charge carriers will produce the mechanical energy change equal to $\Delta W_{je}$ and the dipole moments $\vp_i$ and $\vm_i$ in $\Delta V$ that, along with the induced surface charge in the perfect conductor, generate $\ve^{\rm ins}-\vE^{\rm ins}$ and $\vb^{\rm ins}-\vB^{\rm ins}$.  The total energy supplied by the external source will be the sum of the mechanical energy change in the charge carriers and the energy stored in the fields, that is, the energy density given on the right-hand side of (\ref{44''}).  Since no energy can pass through the perfect conductor enclosing $\Delta V$, the right-hand side of (\ref{44''}) must be greater than or equal to zero in passive material where no energy can be extracted from any initially existing dipole moments.}
\par
A material that satisfies the inequality in (\ref{44up}) can be referred to as being ``unconditionally passive".
An example of a dielectric material satisfying (\ref{44up}), so that the material is unconditionally passive, is one in which its electric polarization is produced by an initially random distribution of isolated electric dipoles that can be modeled by equal and opposite electric charges at the ends of rigid rods with each rigid rod connected to a rigid lattice by an untorqued torsion spring that can be linear or nonlinear, lossy or lossless.
\par
In contrast, we will show that ``permanent" Amperian magnetic dipoles are not unconditionally passive because their initial magnetic dipole moments can be reduced as they align in an external magnetic field.
To illustrate this, consider a causal paramagnetic material with a constant real magnetic susceptibility $\chi_m >0$ over the baseband operational bandwidth $|\w|<\w_0$ such that $\vM = \chi_m \vB /[\mu_0(1+\chi_m)] = \chi_b \vB/\mu_0$, so that $\chi_b>0$ over the baseband bandwidth.   Then, for negligible $\vP$ and the macroscopic fields zero at time $t_0$, the left-hand side of (\ref{44up}), the change in macroscopic polarization energy density, becomes
\be{47}
-\frac{\chi_b}{2\mu_0}|\vB(\vrp,t)|^2 dV < 0
\ee
which contradicts (\ref{44up}).
This contradictory result indicates that the energy change in (\ref{subv6}) and (\ref{44up}) applied to magnetic materials may be a nonnegative energy only for diamagnetic materials, a conclusion proven in Section \ref{mpm} below.
%
\subsection{\label{mpe}{Evaluation of the macroscopic polarization energy for electric dipoles}}
Nearly all electric polarization can be modeled by either initially randomly oriented molecules with permanent electric dipole moments that stay practically fixed in magnitude (like rigid-rod dipoles) as they align in the applied field, or by initially zero electric dipole-moment molecules that distort in the applied field to produce induced electric dipole moments \cite[p. 464]{Kittel}, \cite[p. 162]{Jackson}.  In the latter case, $\Delta W_C(\vrp,t_0)=0$ and the electric dipoles have no initial energy of formation that can be reduced.  In the former case, the energy of formation of the electric dipoles cannot be reduced because the initial separation distance between the equal and opposite charges of each dipole cannot decrease to release part of their initial energy of formation.  (The difference between this former case and the latter case, where the change of energy is manifestly greater than or equal to zero because $\Delta W_C(\vrp,t_0)=0$ and both $\Delta W_C(\vrp,t)$ and $\Delta W_{j_ee}(\vrp,t)$ are $\ge 0$, is that the baseline energy of formation of the electric dipoles in the former case is not zero.  However, since this baseline energy of formation cannot be reduced, the change of energy in this former case of fixed-magnitude electric dipoles is also greater than or equal to zero, as in the latter case.)
\par 
In other words, the change in macroscopic polarization energy for both these models of electric dipoles in passive material satisfies the positive semi-definite inequality in (\ref{44up}). Felsen and Marcuvitz \cite[sec. 1.5]{F&M} arrived at a similar conclusion for a linear, lossless, dispersive, anisotropic medium but without the proviso requiring unconditional passivity, which we shall see is crucially important when we deal with paramagnetic material in the next section.  The derivation of the left-hand side of (\ref{44up}) in terms of the microscopic fields on the right-hand side of (\ref{44up}) has revealed the necessity of assuming unconditional passivity for the microscopic dipoles to ensure that this energy is positive semi-definite.
\par
In summary, assuming either of these two reasonable models for the microscopic electric dipoles in passive material, we find that if the macroscopic magnetization $\vM(\vrp,t)$ in (\ref{44up}) is equal to zero, then
\be{48}
\int\limits_{t_0}^t \frac{\partial\vP(\vrp,t')}{\partial t'}\cdot\vE(\vrp,t') \,
 dt' \approx  U_e(\vrp,t) - U_e(\vrp,t_0) = \frac{1}{\Delta V}\left[\Delta W_{j_ee}(\vrp,t) +\Delta W_C(\vrp,t)-\Delta W_C(\vrp,t_0)\right]
\ge 0\,.
\ee
It is difficult to conceive of a simple model of electric dipoles that would disobey the nonnegative inequality in (\ref{48}).  For example, if the rigid rods separating the equal and opposite electric dipole charges were replaced by initially compressed springs, then the potential energy in the springs could be released upon being uncompressed by applied fields to produce a negative $\Delta W_{j_ee}(t)$ (so the material is nominally nonpassive).
However, the accompanying increase in the formation energy of the electric dipoles caused by the increased separation distance between their equal and opposite charges is greater than the energy released by the springs and, thus,  the associated change in the capacitive energy, $\Delta W_C(\vrp,t) - \Delta W_C(\vrp,t_0)$, would add to $\Delta W_{j_ee}(t)$ to keep (\ref{48}) nonnegative. Conversely, if the springs were compressed further than their initial compression, the decrease in formation energy of the electric dipoles (producing a decrease in $\Delta W_C(\vrp,t) - \Delta W_C(\vrp,t_0)$) would be more than cancelled by the increase in potential energy of the springs (increase in $\Delta W_{j_ee}(t)$)  to, again, keep (\ref{48}) nonnegative.  It should be noted that even for passive material (that is, $\Delta W_{j_ee}(t)\ge 0$) comprised of permanent dipoles with fixed magnitudes (such as the rigid-rod model of electric dipoles), $\Delta W_C(\vrp,t) -\Delta W_C(\vrp,t_0)$ is not necessarily greater than or equal to zero even though $U_e(\vrp,t) - U_e(\vrp,t_0)\ge0$.
\subsection{\label{mpm}{Evaluation of the macroscopic polarization energy for magnetic dipoles}}
Because magnetic charge does not exist, magnetic dipoles are created by molecules with circulating electric-charge currents that can be modeled by perfectly electrically conducting (PEC)\footnote{\lbl{foot8a}Since perfect electric conductors (PEC's) with { infinite conductivity and} zero internal electric and magnetic fields at all frequencies including $\w=0$ do not exist naturally except at superconducting temperatures, it would be more accurate to use the term ``superconductor" rather than ``PEC".  Nevertheless, we will use the classical term ``PEC" in the ideal sense of a ``superconductor".}  wire loops.  If the wire loops carry no permanent current so that all the circulating current and magnetic dipole moments are induced by the applied fields, then the distribution of molecules forms a diamagnetic macroscopic continuum whose low-frequency magnetic susceptibility is less than zero.\footnote{\lbl{foot9}The Bohr-Leeuwen theorem, which states that a classical material composed of free charges in thermal equilibrium cannot be affected by a magnetic field because the magnetic field does no work on moving charges \cite[secs. 24--27]{Vleck}, is often used to argue that classical electromagnetics cannot describe diamagnetism. This theorem assumes that all the energy at thermal equilibrium resides in the kinetic and potential energy of the charge carriers. However, by using a conducting loop model for molecules or metamaterial inclusions, an additional inductive magnetic-field energy is introduced such that the Bohr-Leeuwen theorem no longer applies and we are able to classically derive macroscopic power energy relations in diamagnetic materials and metamaterials. In fact, W. Weber and, to a greater extent, Maxwell, in his Treatise \cite[arts. 836--845]{Maxwell}, explained both diamagnetism and ordinary magnetism by means of the perfectly conducting wire loops  with no ``primitive current" (Maxwell's term for ``permanent current") in the case of diamagnetism, and predominantly ``primitive current" in the case of paramagnetism.}  If the wire loops carry ``permanent" currents with magnetic dipole moments that align in an external magnetic field and dominate any induced diamagnetization, then the distribution of molecules forms a paramagnetic (which includes ferro(i)magnetic and antiferromagnetic \cite[chs. 11 and 12]{Kittel}) macroscopic continuum whose magnetic susceptibility is greater than zero at low frequencies.
\subsubsection{\label{diamagnetism}{Macroscopic polarization energy for diamagnetism}}
In the case of diamagnetism, the initial value of the currents and magnetic dipole moments of the wire loops are zero and $\Delta W_L(\vrp,t_0)=0$. Therefore, passive diamagnetic-material continua that can also contain electric polarization produced by electric dipoles with fixed-magnitude or initially zero electric dipole moments exhibit the unconditional passivity  inequality in (\ref{44up}), specifically
\be{49}
\int\limits_{t_0}^t \left[\frac{\partial\vP(\vrp,t')}{\partial t'}\cdot\vE(\vrp,t')
- \frac{\partial\vB(\vrp,t')}{\partial t'}\cdot\vM(\vrp,t') \right] dt' \ge 0\,.
\ee
As a simple example, consider a medium with negligible loss and constant real permittivity $\eps$ and  permeability $\mu$ over the operational baseband bandwidth $|\w|<\w_0$.  Then $\vP = (\eps-\eps_0)\vE$ and $\vM = (1/\mu_0 -1/\mu)\vB$ over the baseband bandwidth, and (\ref{49}) reduces to
\be{50}
(\eps-\eps_0)|\vE|^2 -(1/\mu_0 -1/\mu)|\vB|^2 \ge 0\,.
\ee
Since $\vE$ and $\vB$ can be chosen equal to zero independently,  this inequality reveals that
\be{51}
\eps \ge \eps_0\,,\;\;\;\; 0\le\mu \le \mu_0
\ee
which confirms that the inequality in (\ref{49}) applies to diamagnetic continua.
\par
The inequality in (\ref{49}) was used in \cite{Yaghjian-Maci-Cloaks}, without showing how it was derived, to obtain  inequalities satisfied by frequency-domain diamagnetic permeability.  A derivation of (\ref{49}) was first given in \cite{Yaghjian-Charleston}.  Inserting $\vP$ and $\vM$ from the constitutive relations (\ref{11'}) into (\ref{49}) recasts this inequality into the form
\be{52}
\int\limits_{t_0}^t \left[\frac{\partial\vD(\vrp,t')}{\partial t'}\cdot\vE(\vrp,t')
+ \frac{\partial\vB(\vrp,t')}{\partial t'}\cdot\vH(\vrp,t') \right] dt' \ge \frac{1}{2}\left[\eps_0 |\vE(\vrp,t)|^2 +\frac{1}{\mu_0}|\vB(\vrp,t)|^2\right]
\ee
which holds for unconditionally passive, spatially nondispersive continua with electric dipolarization and diamagnetization.  An important feature of (\ref{52}) is that the stored macroscopic diamagnetic field energy density is $|\vB(\vrp,t)|^2/(2\mu_0)$, which is not generally equal to $\mu_0|\vH(\vrp,t)|^2/2$ (except in free space).
\subsubsection{\label{paramagnetism}{Macroscopic polarization energy for paramagnetism}}
For paramagnetic substances, which include ferro(i)magnetic and antiferromagnetic substances  \cite[chs. 11 and 12]{Kittel}, the molecules or inclusions contain ``permanent" magnetic dipole moments that can be modeled by randomly oriented, translationally stationary, PEC  wire loops carrying initial currents $I_0$ with magnetic dipole moments $\vm_0$ in the absence of externally applied fields such that when an external field is applied, the magnetization produced by the partial alignment of these magnetic dipole moments dominates the magnetization produced by the induced diamagnetic currents.  The physics would appear to be similar to that of fixed-magnitude permanent electric dipoles and, thus, one is tempted to apply the argument used in Section \ref{mpe} for permanent electric dipoles to conclude that (\ref{49}) and (\ref{52}) hold also for paramagnetic materials.  However, there is an important subtle difference with these ``permanent" Amperian magnetic dipoles that requires a more involved analysis than for the fixed-magnitude permanent electric dipoles.  Even though the ``permanent" magnetic dipole moments can be assumed to dominate over the additional diamagnetic dipole moments induced by applied fields, these small additional induced diamagnetic dipole moments can significantly change (and, in particular, reduce) the initial energy stored in the magnetic fields of the wire loops, as we shall show.  This reduction in initial internal energy invalidates the unconditionally passivity  argument that was used in Section \ref{mpe} for fixed-magnitude electric dipoles to show that (\ref{49}) holds for all $t\ge t_0$.  In fact, as indicated in (\ref{47}), the inequality in (\ref{49}) does not generally hold for paramagnetic continua.
\par
Even if we return to (\ref{39}) and rewrite the last equation in (\ref{39}) as
\bea{39p}
\int\limits_{\Delta V} \left[\frac{\partial\vD(\vrp,t)}{\partial t}\cdot\vE(\vrp,t)
+ \frac{\partial\vB(\vrp,t)}{\partial t}\cdot\vH(\vrp,t) \right] dV
&=& \int\limits_{\Delta V} \left[\frac{\partial\vP(\vrp,t)}{\partial t}\cdot\vE(\vrp,t)
+ \mu_0\frac{\partial\vM(\vrp,t)}{\partial t}\cdot\vH(\vrp,t) \right] dV\nonumber\\ 
&& \hspace{5mm}+ \, \frac{1}{2}\frac{d}{dt}\int\limits_{\Delta V}\left[\eps_0|\vE(\vrp,t)|^2 +\mu_0|\vH(\vrp,t)|^2\right] dV
\eea
and obtain, instead of (\ref{39'}), the equation
\bea{39p'}
\int\limits_{\Delta V} \left[\frac{\partial\vP(\vrp,t)}{\partial t}\cdot\vE(\vrp,t)
+ \mu_0\frac{\partial\vM(\vrp,t)}{\partial t}\cdot\vH(\vrp,t) \right] dV \hspace{50mm}\nonumber\\
\approx \int\limits_{\Delta V} \vj(\vrp,t)\cdot\ve(\vrp,t) dV 
+ \frac{1}{2}\frac{d}{dt}\int\limits_{\Delta V}\left\{\eps_0\left[|\ve(\vrp,t)|^2 -{|}\vE(\vrp,t)|^2\right] +\mu_0\left[|\vh(\vrp,t)|^2 - |\vH(\vrp,t)|^2\right]\right\} dV 
\eea
which, when integrated over time gives
\bea{39p''}
\int\limits_{t_0}^t\int\limits_{\Delta V} \left[\frac{\partial\vP(\vrp,t')}{\partial t'}\cdot\vE(\vrp,t')
+ \mu_0\frac{\partial\vM(\vrp,t')}{\partial t'}\cdot\vH(\vrp,t') \right] dV dt' \hspace{50mm}\nonumber\\
\approx \int\limits_{t_0}^t\int\limits_{\Delta V} \vj(\vrp,t')\cdot\ve(\vrp,t') dV dt'
+ \frac{1}{2}\int\limits_{\Delta V}\left\{\eps_0\left[|\ve(\vrp,t')|^2 -{|}\vE(\vrp,t')|^2\right] +\mu_0\left[|\vh(\vrp,t')|^2 - |\vH(\vrp,t')|^2\right]\right\}_{t_0}^t dV 
\eea
it is not apparent that the right-hand side of this equation in (\ref{39p''}) is always greater than or equal to zero for paramagnetic macroscopic continua because the initial energy of formation of the Amperian magnetic dipoles can be reduced.  Also note in (\ref{39p''}) that $|\vh|^2 - |\vH|^2 \neq |\vh-\vH|^2$ because for Amperian magnetic dipoles $\int_{\Delta V}\vh dV = \int_{\Delta V}\vb dV /\mu_0 =\vB\Delta V/\mu_0 \neq \vH \Delta V$.
\par
To derive a positive semi-definite macroscopic polarization energy for the ``permanent" microscopic Amperian magnetic dipoles of paramagnetic materials, we begin by evaluating $P_{je}(t)$ for a microscopic (electrically small), perfectly electrically conducting, thin, rigid, wire loop spanning the area $A$ and carrying an initial current $I_0$ such that the initial magnetic dipole moment of the loop is $\vm_0 = I_0 A\vnh$, where $\vnh$ is the normal to the loop in the direction determined by the right-hand rule applied to the circulation of $I_0$.  For the sake of simplicity, assume that the thin wire loop lies in a plane (see Fig. \ref{fig3}), although the derivation goes through for electrically small wire loops of any shape characterized by a self-inductance.  The loop can be considered bound to a rigid lattice by lossless or lossy torsion springs that allow the loop to rotate about axes through its center $\vrp_0$.  With the wire loop illuminated by external electric and magnetic fields,  $\ve_{\rm ext}(\vrp,t)$ and $\vb_{\rm ext}(\vrp,t)$, we can write 
\be{P1}
P_{j_be}(t) = \int\limits_V \vj_b(\vrp,t)\cdot\ve(\vrp,t) dV = \int\limits_V \vj_b(\vrp,t)\cdot\ve_{\rm ext}(\vrp,t) dV + \int\limits_V \vj_b(\vrp,t)\cdot\ve_{\rm ind}(\vrp,t) dV
\ee
where $\ve_{\rm ind}(\vrp,t)$ is the induced electric field, that is, the electric field generated by the current $\vj_b(\vrp,t)$ in the loop.  Note that because the loop can rotate, the total electric field $\ve =\ve_{\rm ext}+\ve_{\rm ind}$ is not equal to zero in the PEC wire.  Rather it is the Lorentzian force field that is zero in the rotating PEC wire, that is
\be{P2}
\ve(\vrp,t) +\vv(\vrp,t)\times\vb(\vrp,t) =0
\ee
where $\vv(\vrp,t)$ is the velocity of the point $\vrp$ in the wire at the time $t$ and the total magnetic field can be written as the sum of the externally applied magnetic field and the induced magnetic field, $\vb =\vb_{\rm ext}+\vb_{\rm ind}$.  The induced magnetic field $\vb_{\rm ind}(\vrp,t)$ is produced by the current in the loop, including the initial current $I_0$.
\begin{figure}[ht]
\begin{center}
\includegraphics[width =4.0in]{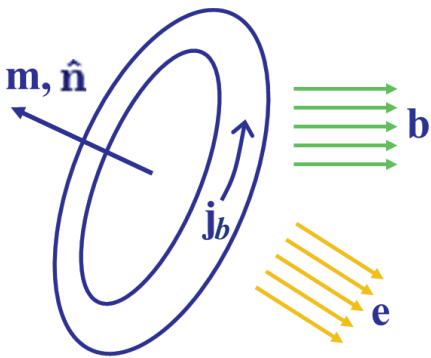}
\end{center}
\caption{\label{fig3}Perfectly electrically conducting (PEC), electrically small, rotating, thin, rigid wire loop carrying an initial current $I_0$ and subject to an external field so that the total microscopic current distribution is $\vj_b$, which produces a magnetic dipole moment $\vm$.   The total microscopic electric and magnetic fields, $\ve$ and $\vb$,  are equal to the sums of the external fields and the induced fields produced by $\vj_b$.}
\end{figure}
%
\par
For an electrically small loop carrying a current 
$I(t)$ at the time $t$, the integral in (\ref{P1}) involving $\ve_{\rm ext}$ can be evaluated quasi-statically as
\bea{P4}
\int\limits_V \vj_b(\vrp,t)\cdot\ve_{\rm ext}(\vrp,t) dV = I(t)\oint\limits_{C(t)}\ve_{\rm ext}(\vrp,t) \cdot d\vc = -I(t)\int\limits_{S(t)}\frac{\partial\vb_{\rm ext}}{\partial t}(\vrp,t) \cdot \vnh dS\nonumber\\
= -I(t)\frac{\partial\vb_{\rm ext}(\vrp_0,t)}{\partial t} \cdot\int\limits_S \vnh dS = -I(t) A\vnh\cdot\frac{\partial\vb_{\rm ext}(\vrp_0,t)}{\partial t}  = -\vm(t)\cdot\frac{\partial\vb_{\rm ext}(\vrp_0,t)}{\partial t} 
\eea
where $C(t)$ and $S(t)$ are the rotating curve and planar surface, respectively, defining the thin wire loop at the time $t$, and $\vch$ is the unit vector along the direction of the wire.  The result in (\ref{P4}) expresses the power supplied by the externally applied voltage to the current in the PEC wire loop in terms of the magnetic dipole moment and the time derivative of the external magnetic field at each instant of time $t$.
\par
A relatively easy way to evaluate the integral in (\ref{P1}) involving $\ve_{\rm ind}$ for an electrically small loop is to extend the integration from $V$ to all space $V_\infty$, that is
\be{P5}
\int\limits_V \vj_b(\vrp,t)\cdot\ve_{\rm ind}(\vrp,t) dV = \int\limits_{V_\infty} \vj_b(\vrp,t)\cdot\ve_{\rm ind}(\vrp,t) dV.
\ee
Then, insert $\vj_b$ from Ampere's law (Maxwell's second equation), $\vj_b = \nabla\times\vb_{\rm ind}/\mu_0 -\eps_0\partial \ve_{\rm ind}/\partial t$, to get
\be{P6}
\int\limits_{V_\infty} \vj_b(\vrp,t)\cdot\ve_{\rm ind}(\vrp,t) dV = \int\limits_{V_\infty}\left(\frac{1}{\mu_0}\nabla\times\vb_{\rm ind} -\eps_0\frac{\partial \ve_{\rm ind}}{\partial t}\right)\cdot\ve_{\rm ind} dV
\ee
which, with the help of the equations, $\nabla\cdot(\ve_{\rm ind}\times\vb_{\rm ind}) = (\nabla\times\ve_{\rm ind})\cdot\vb_{\rm ind}-(\nabla\times\vb_{\rm ind})\cdot\ve_{\rm ind}$ and $\nabla\times\ve_{\rm ind}= - \partial \vb_{\rm ind}/\partial t$, converts to 
\be{P7}
\int\limits_{V_\infty} \vj_b(\vrp,t)\cdot\ve_{\rm ind}(\vrp,t) dV = -\frac{1}{2}\int\limits_{V_\infty}\frac{\partial}{\partial t}\left(\frac{1}{\mu_0}|\vb_{\rm ind}|^2 + \eps_0|\ve_{\rm ind}|^2\right) dV
= -\frac{1}{2}\frac{d}{dt}\int\limits_{V_\infty}\left(\frac{1}{\mu_0}|\vb_{\rm ind}|^2 + \eps_0|\ve_{\rm ind}|^2\right) dV
\ee
after noting that the integral over $S_\infty$ of the quasi-static Poynting vector is zero.  The energy in the quasi-electrostatic field of the inductor is negligible compared to the energy in its quasi-magnetostatic field and, thus, (\ref{P7}) can be rewritten as
\be{P11}
\int\limits_V \vj_b(\vrp,t)\cdot\ve_{\rm ind}(\vrp,t) dV =-\frac{d}{dt}\left[\frac{1}{2\mu_0}\int\limits_{V_\infty}|\vb_{\rm ind}(\vrp,t)|^2 dV\right] = -\frac{d}{dt}\left[\frac{1}{2}L I^2(t)\right]
\ee
where $L$ is the inductance of the PEC wire loop \cite[ch. 8]{Plonsey&Collin}.
\par
With the results in (\ref{P11}) and (\ref{P4}) inserted into (\ref{P1}), we find that the total power supplied by the fields to the current loop is given by
\be{P12}
P_{j_be}(t) = \int\limits_V \vj_b(\vrp,t)\cdot\ve(\vrp,t) dV =  -\vm(t)\cdot\frac{\partial\vb_{\rm ext}(\vrp_0,t)}{\partial t}  - \frac{d}{dt}\left[\frac{1}{2}L I^2(t)\right].
\ee
This equation (\ref{P12}) says that the power $P_{j_be}(t)$ supplied by the electric field to the current carrying PEC wire loop is equal to the total power supplied by the external fields minus the time rate of change of the energy stored in the quasi-magnetostatic field of the current loop.  As explained in Section \ref{Energy}, the power $P_{j_be}(t)$ in (\ref{P12}) is also equal to the time rate of change of the mechanical energy, that is, the sum of (i) the kinetic energy of the charge carriers and the kinetic energy of the rotating PEC conductors, (ii) the kinetic and potential energy of the torsion springs attaching the rotating loop to the rigid lattice, and (iii) any heat loss in the torsion springs.
\par
The inductance power on the right-hand side of (\ref{P12}) can also be expressed in terms of the magnetic dipole moment $\vm$ and the externally applied magnetic field $\vb_{\rm ext}$ by invoking the circuit-theory result (from the quasi-static Maxwell equations) that the electromotive force applied around the electrically small rotating PEC wire-loop circuit by the externally applied fields equals the product of the inductance and the time derivative of the current, that is 
\be{P13}
\oint\limits_{C(t)}[\ve_{\rm ext}(\vrp,t) + \vv(\vrp,t)\times\vb_{\rm ext}(\vrp,t)]\cdot d\vc = L \frac{dI}{dt}
\ee
where $\vv(\vrp,t)$ is the velocity of the thin rotating loop (defined by the closed curve $C(t)$) at each point $\vrp$ of $C(t)$ at the time $t$.  With the aid of Faraday's law for a moving curve \cite[pp. 39--40]{A&B}, namely
\be{P14}
\oint\limits_{C(t)}[\ve_{\rm ext}(\vrp,t)+\vv(\vrp,t)\times\vb_{\rm ext}(\vrp,t)] \cdot d\vc = -\frac{d}{dt}\int\limits_{S(t)} \vb_{\rm ext}(\vrp,t)\cdot\vnh dS \approx -\frac{d}{dt}\left[\vb_{\rm ext}(\vrp_0,t)\cdot\vnh A\right]
\ee
we have
\be{P14'}
 L \frac{dI}{dt}= -\frac{d}{dt}\int\limits_{S(t)} \vb_{\rm ext}(\vrp,t)\cdot\vnh dS \approx -\frac{d}{dt}\left[\vb_{\rm ext}(\vrp_0,t)\cdot\vnh A\right]
\ee 
and thus
\be{P15}
\frac{d}{dt}\left[\frac{1}{2}L I^2(t)\right]=LI\frac{dI}{dt} = -I\frac{d}{dt}\left[\vb_{\rm ext}(\vrp_0,t)\cdot\vnh A\right] = -\frac{d}{dt}\left[\vb_{\rm ext}(\vrp_0,t)\cdot\vm \right] + \frac{dI}{dt}\left[\vb_{\rm ext}(\vrp_0,t)\cdot\vnh A \right]
\ee
so that (\ref{P12}) becomes
\be{P16}
P_{j_be}(t) = \int\limits_V \vj_b(\vrp,t)\cdot\ve(\vrp,t) dV = \vb_{\rm ext}(\vrp_0,t)\cdot\left(\frac{d\vm}{dt} -\vnh \frac{dm}{dt}\right)
\ee
in which we have used the identity $A\,dI/dt = dm/dt$.  Note that $P_{j_be}(t)$ is zero if the wire is not rotating (because then $\vnh dm/dt =d(m\vnh)/dt = d\vm/dt$), a result that checks with $\ve=0$ in a stationary nonrotating PEC wire loop.  Since we are assuming that we are dealing with paramagnetic  materials in which the induced diamagnetic moments are negligible compared with the aligning (rotating) ``permanent" magnetic dipole moments, it follows that $|\vnh dm/dt|= |dm/dt| \ll |d\vm/dt|\approx m_0|d\vnh/dt|$ and (\ref{P16}) reduces to simply
\be{P17}
P_{j_be}(t) = \int\limits_V \vj_b(\vrp,t)\cdot\ve(\vrp,t) dV \approx m_0\vb_{\rm ext}(\vrp_0,t)\cdot\frac{d\vnh(t)}{dt} \approx \vb_{\rm ext}(\vrp_0,t)\cdot\frac{d\vm(t)}{dt}\,.
\ee
Although the magnitude of the induced magnetic dipole moment is negligible compared to the magnitude of the total magnetic dipole moment as the dipole aligns in an external magnetic field, the change in energy produced by the induced magnetic dipole moment is not negligible.
\par
The result in (\ref{P17}), which was also obtained in \cite[eq. (2.164)]{H&Y}  for ``permanent" current enforced on a PEC sphere\footnote{\lbl{footsphere}Although not stated in \cite{H&Y}, the electrically small PEC sphere with its static dipole moment has to be rotating.} rather than for the ``permanent" current on a PEC wire loop,  is quite revealing.  \textit{It says that the power supplied by the electric field applied to an inclusion or molecule with a ``permanent" Amperian magnetic dipole moment (modeled by the PEC wire loop carrying the initial current $I_0$ with magnetic dipole moment $\vm_0$) that predominates over any induced diamagnetic moment (as is apparently the case for most if not all known natural paramagnetic materials) is the same as the power supplied by the magnetic field to a hypothetical fixed-magnitude ($m_0$) magnetic dipole moment formed by separated equal and opposite magnetic charge} \cite[eq. (2.167)]{H&Y}.\footnote{\lbl{foot10}In particular, we have
\be{fm1}
\int\limits_V \vj_b(\vrp,t)\cdot\ve(\vrp,t) dV = \int\limits_V \vj_m(\vrp,t)\cdot\vh_c(\vrp,t) dV
\ee
where $\vj_m$ is the magnetic current of the rotating fixed-magnitude magnetic-charge magnetic dipole and $\vh_c$ is the total (external plus induced) magnetic field.  (For fixed-magnitude magnetic-charge magnetic dipoles, $\int_V \vj_m \cdot \vh_{c,\rm ind} dV =0$.)  Since the electric and magnetic fields well outside electrically small dipoles are the same whether they are produced by electric or magnetic charge-current, Poynting's theorem applied to a free-space surface enclosing but well away from either an Amperian or magnetic-charge magnetic dipole has the same $\vnh\cdot(\ve\times\vh)=\vnh\cdot(\ve_c\times\vh_c)$ and thus the same free-space stored energy $\int_V(\eps_0|\ve|^2 +|\vb|^2/\mu_0) dV/2 =\int_V(\eps_0|\ve_c|^2 +\mu_0|\vh_c|^2) dV/2 $ because the energy in (\ref{fm1}) supplied to each of the dipoles is also the same.}
\par
Under the same condition that the rotating ``permanent" magnetic dipole moments dominate the induced magnetic dipole moments, we have from (\ref{P15}) and (\ref{P11}) that
\be{P18}
\int\limits_V \vj_b(\vrp,t)\cdot\ve_{\rm ind}(\vrp,t) dV = -\frac{d}{dt}\left[\frac{1}{2}L I^2(t)\right]\approx \frac{d}{dt}\left[\vb_{\rm ext}(\vrp_0,t)\cdot\vm \right]
\ee
which can also be obtained by subtracting (\ref{P4}) from(\ref{P17}).
This power supplied by the internal electric field to the current in the PEC wire loop is the microscopic ``hidden power" corresponding to the time rate of change of hidden momentum  \cite[pp. 214--216, 244]{P&H}, \cite{Suttorp, McDonald, Mansuripur2013} for Amperian magnetic dipoles in the electromagnetic force-momentum equation that is analogous to the power-energy equation (Poynting's theorem).  Specifically, the time rate of change of hidden momentum corresponding to (\ref{P18}) is equal to $-(1/c^2)d(\vm\times\ve_{\rm ext})/dt$ \cite[eq. (2.161)]{H&Y}.  Here, our detailed microscopic derivation has shown that the microscopic ``hidden power" is drawn from the reservoir of inductive energy in the initial microscopic Amperian magnetic dipole moment.  
\par
Referring to natural materials or metamaterials whose magnetism is dominated by the alignment of ``permanent" Amperian magnetic dipoles as \textit{{paramagnetic} materials}, we have shown that {paramagnetic} materials behave {energetically} as if their  magnetic dipole moments were produced by unconditionally passive, fixed-magnitude magnetic-charge dipoles (for example, equal and opposite magnetic charges at the ends of rigid rods).  Although energy can be extracted from the initial internal inductive energy of the wire loops (Amperian magnetic dipoles) to invalidate the unconditional passivity used to derive (\ref{49}), the external and internal power on the right-hand side of (\ref{P12}) combine to form a total power in (\ref{P17}) supplied by the electric field to the Amperian magnetic dipole that is equivalent to the power supplied by the magnetic field to an unconditionally passive, magnetic-charge magnetic dipole with the same initial magnetic dipole moment.  Consequently, a convenient way to determine positive semi-definite energy relations for macroscopic {paramagnetic} dipolar continua that correspond to the energy relations in (\ref{49}) and (\ref{52}) for diamagnetic dipolar continua is to repeat the analysis in Sections \ref{ME} through \ref{mpe} beginning with the {Maxwell} microscopic equations that include microscopic magnetic charge, namely \cite[p. 464]{Stratton} 
\begin{subequations}
\label{M1}
\be{M1a}
\nabla\times\ve_c(\vrp,t) +\mu_0\frac{\partial \vh_c(\vrp,t)}{\partial t} = -\vj_m(\vrp,t)
\ee
\be{M1b}
\nabla\times\vh_c(\vrp,t) -\eps_0\frac{\partial \ve_c(\vrp,t)}{\partial t} = \vj_e(\vrp,t)
\ee
\be{M1c}
\nabla\cdot\vh_c(\vrp,t) =\frac{1}{\mu_0}\varrho_m(\vrp,t)
\ee
\be{M1d}
\eps_0\nabla\cdot\ve_c(\vrp,t) = \varrho_e(\vrp,t)
\ee
\end{subequations}
where $\varrho_m(\vrp,t)$ and $\vj_m(\vrp,t)$ are the magnetic charge and current densities, and $\ve_c(\vrp,t)$ and $\vh_c(\vrp,t)$ are the primary fields in the magnetic-charge magnetic dipole system (with electric-charge electric dipoles).  It is assumed that any electric and magnetic dipole moments are produced by electric and magnetic charge separation ($\varrho_e$ and $\varrho_m$), respectively, and that the electric and magnetic currents, $\vj_e$ and $\vj_m$, do not produce Amperian magnetic dipoles or analogous circulating-magnetic-current electric dipoles, respectively.  Because the power and energy derivations for the {paramagnetic} dipoles proceed parallel to those already given in detail for the electric dipoles, only the main results will be given.
\par
Instead of (\ref{5'}), we have for the microscopic {Poynting} theorem that includes microscopic magnetic charge-current 
\bea{M5'}
P(t) = -\int\limits_S \vnh\cdot[\ve_c(\vrp,t)\times\vh_c(\vrp,t)] dS \hspace{51.5mm}\nonumber\\ =   \int\limits_V [\vj_e(\vrp,t)\cdot\ve_c(\vrp,t)+ \vj_m(\vrp,t)\cdot\vh_c(\vrp,t)] dV   + \frac{1}{2}\frac{d}{dt}\int\limits_V\left[\eps_0|\ve_c(\vrp,t)|^2 +\mu_0|\vh_c(\vrp,t)|^2\right] dV.
\eea
Instead of (\ref{5W}), we have
\be{M5W}
W_{jem}(t) =\int\limits_{t_0}^t \int\limits_V [\vj_e(\vrp,t')\cdot\ve_c(\vrp,t')+ \vj_m(\vrp,t')\cdot\vh_c(\vrp,t')] dV dt' \ge 0
\ee
in a passive material with applied fields zero until after the time $t_0$.  
\par
One can replace $\vh_c$ with $\vb_c/\mu_0$ in (\ref{M1})--(\ref{M5W}) because these equations hold in free space.  However, to derive the conventional form of Maxwell's macroscopic equations in dipolar continua from the microscopic equations in (\ref{M1}), $\vh_c$ must be retained in (\ref{M1}) because now the magnetic dipole moments and magnetization are defined in terms of magnetic charge rather than circulating electric current.  In particular, the equation (\ref{21b}) is replaced by
\be{M21b}
\vM(\vrp,t) = \frac{1}{\mu_0\Delta V}\int\limits_{\Delta V} \varrho_{mb}(\vrp+\vrp',t)\vrp' dV'
\ee
which leads to the conventional form of Maxwell's macroscopic equations in (\ref{12'}) if and only if one begins with $\vh_c$ in (\ref{M1}) and not $\vb_c/\mu_0$.
\par
Since the power in (\ref{M5'}) is still given by the usual integral of the macroscopic Poynting vector in a spatially nondispersive continua, the analysis in Section \ref{PTDC} again leads to (\ref{34}) and (\ref{36}).  However, the analysis in Section \ref{ER} that led to (\ref{44''}), (\ref{49}) and (\ref{52}) for diamagnetic material now leads to the following three analogous macroscopic inequalities{, (\ref{44p''}), (\ref{M49}) and (\ref{M52}),} for paramagnetic material
\be{44p''}
\int\limits_{t_0}^t \left[\frac{\partial\vP(\vrp,t')}{\partial t'}\cdot\vE(\vrp,t')
+ \mu_0\frac{\partial\vM(\vrp,t')}{\partial t'}\cdot\vH(\vrp,t') \right] dt' 
\approx  U_e(\vrp,t) -U_e(\vrp,t_0)+ U_h(\vrp,t) -U_h(\vrp,t_0)
\ee
in which 
\begin{subequations}
\lbl{ped'}
\bea{peda'}
U_e(\vrp,t) =  \frac{1}{\Delta V} \int\limits_{t_0}^t\int\limits_{\Delta V} \vj_e\cdot\ve_c dVdt' 
+ \frac{1}{2\Delta V}\int\limits_{\Delta V}\eps_0|\ve^{\rm ins}(\vrp,t) -\vE^{\rm ins}(\vrp,t)|^2 dV\nonumber\\ 
= \frac{1}{\Delta V}\left[\Delta W_{j_ee_c}(\vrp,t) +\Delta W_C(\vrp,t)\right]
\eea
\bea{pedb'}
U_h(\vrp,t) =  \frac{1}{\Delta V} \int\limits_{t_0}^t\int\limits_{\Delta V} \vj_m\cdot\vh_c dVdt' 
+ \frac{1}{2\Delta V}\int\limits_{\Delta V}\mu_0|\vh_c^{\rm ins}(\vrp,t) - \vH^{\rm ins}(\vrp,t)|^2  dV\nonumber\\
= \frac{1}{\Delta V}\left[\Delta W_{j_mh_c}(\vrp,t) +\Delta W_{Lh}(\vrp,t)\right]
\eea
\end{subequations}
where
\begin{subequations}
\lbl{5WCL''}
\be{5WCL''a}
\Delta W_{jem}(\vrp,t) = \Delta W_{j_ee_c}(\vrp,t)+\Delta W_{j_mh_c}(\vrp,t) = \int\limits_{t_0}^t \int\limits_{\Delta V} [\vj_e(\vrp,t')\cdot\ve_c(\vrp,t')+\vj_m(\vrp,t')\cdot\vh_c(\vrp,t')] dV dt'
\ee
with
\be{5WCL''b}
\Delta W_{j_ee_c}(\vrp,t) = \int\limits_{t_0}^t \int\limits_{\Delta V} \vj_e(\vrp,t')\cdot\ve_c(\vrp,t') dV dt'
\ee
\be{5WCL''c}
\Delta W_{j_mh_c}(\vrp,t) = \int\limits_{t_0}^t \int\limits_{\Delta V} \vj_m(\vrp,t')\cdot\vh_c(\vrp,t') dV dt'.
\ee
\end{subequations}
The averaged macroscopic fields and sources are the same for the Amperian and magnetic-charge magnetic dipoles.
\par
The energy density in (\ref{44p''}) is greater than or equal to zero because the initial energy of formation of the rigid-rod, magnetic-charge magnetic dipoles (like the rigid-rod electric-charge electric dipoles) cannot be reduced to release energy and thus
\be{M49}
\int\limits_{t_0}^t \left[\frac{\partial\vP(\vrp,t')}{\partial t'}\cdot\vE(\vrp,t')
+\mu_0 \frac{\partial\vM(\vrp,t')}{\partial t'}\cdot\vH(\vrp,t') \right] dt' \ge 0
\ee
or
\be{M52}
\int\limits_{t_0}^t \left[\frac{\partial\vD(\vrp,t')}{\partial t'}\cdot\vE(\vrp,t')
+ \frac{\partial\vB(\vrp,t')}{\partial t'}\cdot\vH(\vrp,t') \right] dt' \ge \frac{1}{2}\left[\eps_0 |\vE(\vrp,t)|^2 +\mu_0|\vH(\vrp,t)|^2\right]
\ee
in paramagnetically unconditionally passive (with respect to the rigid-rod, magnetic-charge formulation of magnetic dipoles), spatially nondispersive material whose applied fields are zero until after the initial time $t_0$.  The two energy equations in (\ref{M49}) and (\ref{M52}) were first obtained in \cite{Y&B, Yaghjian-Felsen} but without the rigor of the foregoing derivation of these macroscopic equations from the microscopic equations.  Also, in \cite{Y&B, Yaghjian-Felsen} it was not explicitly pointed out that these equations do not necessarily apply to diamagnetic continua or to spatially dispersive continua with $\vnh\cdot(\vE\times\vH)$ discontinuous across a free-space/material interface; see Section \ref{PTDC}.
\section{\label{RER}{RECAPITULATION OF ENERGY RELATIONS FOR MACRO\-SCO\-PIC DIPOLAR CONTINUA}}
Using the macroscopic Maxwell equations and constitutive relations in (\ref{12'}) and (\ref{11'}), we can rewrite the 
energy relations for unconditionally passive, spatially nondispersive (more generally, continuous $\vnh\cdot(\vE\times\vH)$ across interfaces), electric-dipole/diamagnetic continua as
\bea{49'}
\int\limits_V\int\limits_{t_0}^t \left[\frac{\partial\vP(\vrp,t')}{\partial t'}+\nabla'\times \vM(\vrp,t') \right]\cdot\vE(\vrp,t') dt' dV\hspace{5cm}\nonumber\\ =
\int\limits_V\int\limits_{t_0}^t \left[\frac{\partial\vP(\vrp,t')}{\partial t'}\cdot\vE(\vrp,t')
- \frac{\partial\vB(\vrp,t')}{\partial t'}\cdot\vM(\vrp,t') \right] dt' dV \ge 0
\eea
or
\bea{52'}
-\int\limits_S\int\limits_{t_0}^t \vnh\cdot[\vE(\vrp,t')\times\vH(\vrp,t')]dt' dS=\int\limits_V\int\limits_{t_0}^t \left[\frac{\partial\vD(\vrp,t')}{\partial t'}\cdot\vE(\vrp,t')
+ \frac{\partial\vB(\vrp,t')}{\partial t'}\cdot\vH(\vrp,t') \right] dt'dV\nonumber\\ \ge \frac{1}{2}\int\limits_V\left[\eps_0 |\vE(\vrp,t)|^2 +\frac{1}{\mu_0}|\vB(\vrp,t)|^2\right]dV \hspace{1.5cm}
\eea
for all volumes $V$ with macroscopic fields zero until after the time $t_0$.  For the first equation in (\ref{49'}) to hold, the surface $S$ of the volume $V$ in (\ref{49'}) must lie in a thin free-space shell enclosing the volume $V$ such that the equivalent electric surface currents (delta functions in $\nabla'\times \vM(\vrp,t')$) are included in the volume integration; see Section \ref{PTDC}.  (In (\ref{52'}) it is irrelevant whether or not $S$ lies in free space because $\vnh\cdot(\vE\times\vH)$ for the material being considered is continuous across a free-space/material interface.)  Equation (\ref{49'}) confirms that the equivalent electric volume-current power density, $\vJ_e^{\rm eq}(\vrp,t)\cdot\vE(\vrp,t)$, in Maxwell's macroscopic equations for electric-dipole/diamagnetic material is given by $[\partial \vP(\vrp,t)/\partial t + \nabla\times \vM(\vrp,t)]\cdot \vE(\vrp,t)$, that is, $\vJ_e^{\rm eq}=\partial \vP/\partial t + \nabla\times \vM$.  Recall that diamagnetic material is defined herein as material with magnetization produced by molecules or inclusions that have no permanent magnetic dipole moments, only induced magnetic dipole moments.  Equations (\ref{49'})--(\ref{52'}) hold for dipolar metamaterials whose inclusions have no permanent dipole moments as long as the frequency is low enough that these metamaterials behave as spatially nondispersive dipolar continua; for example, (\ref{49'})--(\ref{52'}) apply to spatially nondispersive bulk metamaterial continua made of split resonant rings \cite{Marques}.
\par
Similarly, for unconditionally passive (with respect to equivalent hypothetical magnetic-charge magnetic dipoles), spatially nondispersive (more generally, continuous $\vnh\cdot(\vE\times\vH)$ across interfaces), electric-dipole/paramagnetic (including ferro(i)magnetic and antiferromagnetic\footnote{\lbl{footanti}Antiferromagnetic materials are comprised of permanent magnetic dipoles that align antiparallel to each other below the N\'eel temperature to produce a paramagnetism in an applied field with small positive-susceptibility, macroscopically-averaged magnetization and fields (even though their initial microscopic fields and energies can be large) \cite[ch. 12]{Kittel}.  Thus, the macroscopic fields of an\-ti\-fer\-ro\-mag\-ne\-tic materials satisfy the paramagnetic inequalities in (\ref{49''})--(\ref{52''}). }) continua
\be{49''}
\int\limits_V\int\limits_{t_0}^t \left[\frac{\partial\vP(\vrp,t')}{\partial t'}\cdot\vE(\vrp,t')+\mu_0\frac{\partial\vM(\vrp,t')}{\partial t'}\cdot \vH(\vrp,t') \right] dt' dV \ge 0
\ee
or
\bea{52''}
-\int\limits_S\int\limits_{t_0}^t \vnh\cdot[\vE(\vrp,t')\times\vH(\vrp,t')]dt' dS=\int\limits_V\int\limits_{t_0}^t \left[\frac{\partial\vD(\vrp,t')}{\partial t'}\cdot\vE(\vrp,t')
+ \frac{\partial\vB(\vrp,t')}{\partial t'}\cdot\vH(\vrp,t') \right] dt'dV\nonumber\\ \ge \frac{1}{2}\int\limits_V\left[\eps_0 |\vE(\vrp,t)|^2 +\mu_0|\vH(\vrp,t)|^2\right]dV \hspace{1.5cm}
\eea
for all volumes $V$ with macroscopic fields zero until after the time $t_0$.  In either of these equations, it is irrelevant whether or not the surface $S$ of the volume $V$  lies in a thin free-space shell enclosing the volume $V$.
 Equation (\ref{49''}) confirms that the equivalent electric-plus-magnetic volume-current power density, $[\vJ_e^{\rm eq}(\vrp,t)\cdot\vE(\vrp,t)+\vJ^{\rm eq}_m(\vrp,t)\cdot\vH(\vrp,t)]$, in Maxwell's macroscopic equations for electric-dipole/{paramagnetic} material is given by $[\partial \vP(\vrp,t)/\partial t \cdot \vE(\vrp,t) + \mu_0\partial \vM(\vrp,t)/\partial t \cdot \vH(\vrp,t))]$, that is, $\vJ_e^{\rm eq}=\partial \vP/\partial t$ and $\vJ_m^{\rm eq}=\mu_0\partial \vM/\partial t$.  Recall that {paramagnetic} material is defined herein as material with magnetization produced by the alignment of ``permanent" magnetic dipole moments of molecules or inclusions that dominate any induced diamagnetic magnetization.  Metamaterials with inclusions characterized by {paramagnetic $\vm$ producing a macroscopic paramagnetic magnetization $\vM = \int_{\Delta V}\vrp\times(\nabla\times\vm)dV/(2\Delta V) = \int_{\Delta V}\vm dV/\Delta V$ that dominates over any diamagnetism produced by the inclusion's conduction and electric polarization currents} would have macroscopic fields satisfying the inequalities in (\ref{49''})--(\ref{52''}). 
\par
The difference between the power density integrands in the {paramagnetic} and diamagnetic continua in (\ref{49''}) and (\ref{49'}), respectively, is equal to the macroscopic ``hidden power"
\begin{subequations}
\label{HP}
\be{HPa}
\frac{\partial}{\partial t}\left(\vM\cdot\vB - \frac{\mu_0}{2}|\vM|^2\right) =\frac{1}{2}\frac{\partial}{\partial t}\left(\frac{1}{\mu_0}|\vB|^2 -\mu_0|\vH|^2\right)
\ee
with the associated macroscopic ``hidden energy"
\be{HPb}
\left(\vM\cdot\vB - \frac{\mu_0}{2}|\vM|^2\right) =\frac{1}{2}\left(\frac{1}{\mu_0}|\vB|^2 -\mu_0|\vH|^2\right)
\ee
\end{subequations}
analogous to the ``hidden momentum" between the Abraham and Minkowski formulations of the macroscopic electromagnetic force-momentum equation \cite[pp. 214--216, 244]{P&H},  \cite{Suttorp, McDonald, Mansuripur2013}.  As mentioned above, our detailed microscopic derivation has shown that the macroscopic ``hidden energy" is drawn from the reservoir of inductive energy in the initial paramagnetic microscopic Amperian magnetic dipole moments.   {For fields that become periodic after $t=t_0$, the time-average of the hidden energy in (\ref{HPb}) approaches a constant.}
\par
The positive semi-definite inequalities of the equations in (\ref{49'})--(\ref{52'}) and (\ref{49''})--(\ref{52''}) were derived from the microscopic equations and thus they have not required linearity or any specific constitutive equations.  The functional dependence of the macroscopic polarizations on the fields can be nonlinear and anisotropic (or bianisotropic).  Both sets of equations, (\ref{49'})--(\ref{52'}) for electric-dipole/diamagnetic macroscopic continua and (\ref{49''})--(\ref{52''}) for electric-dipole/paramagnetic macroscopic continua, hold to a more accurate approximation the better the macroscopic continua approximate ideal continua, that is, the smaller the values of $k_0^{\rm max}d$ and $k^{\rm max}d$ are than unity.  Also, a corollary of both these sets of equations is that
\be{52cor}
-\int\limits_S\int\limits_{t_0}^t \vnh\cdot[\vE(\vrp,t')\times\vH(\vrp,t')]dt' dS=\int\limits_V\int\limits_{t_0}^t \left[\frac{\partial\vD(\vrp,t')}{\partial t'}\cdot\vE(\vrp,t') 
+ \frac{\partial\vB(\vrp,t')}{\partial t'}\cdot\vH(\vrp,t') \right] dt'dV \ge 0
\ee
for both diamagnetic and paramagnetic continua.  It follows from (\ref{52cor}), (\ref{33}), and (\ref{5'}) that
\bea{5''}
-\int\limits_S\int\limits_{t_0}^t \vnh\cdot[\ve(\vrp,t')\times\vh(\vrp,t')] dt'dS  = -\int\limits_S\int\limits_{t_0}^t \vnh\cdot[\ve_c(\vrp,t')\times\vh_c(\vrp,t')]dt' dS \hspace{20mm}\nonumber\\ =   \int\limits_V \int\limits_{t_0}^t\vj(\vrp,t')\cdot\ve(\vrp,t')dt' dV   + \frac{1}{2}\int\limits_V\left[\eps_0|\ve(\vrp,t')|^2 +|\vb(\vrp,t')|^2/\mu_0\right]_{t_0}^t dV\hspace{10mm}\\ = \int\limits_V \int\limits_{t_0}^t[\vj_e(\vrp,t')\cdot\ve_c(\vrp,t') +\vj_m(\vrp,t')\cdot\vh_c(\vrp,t')] dt'dV   + \frac{1}{2}\int\limits_V\left[\eps_0|\ve_c(\vrp,t')|^2 
+\mu_0|\vh_c(\vrp,t')|^2\right]_{t_0}^t dV \ge 0\,.\nonumber
\eea
Thus, both the total microscopic and macroscopic electromagnetic energy entering an initially unexcited, passive, spatially nondispersive volume of either diamagnetic or paramagnetic dipolar material is always nonnegative provided the surface of the volume lies in free space and does not cut through the dipoles.  It also follows from (\ref{49''}) that (\ref{39p''}) is greater than or equal to zero for paramagnetic continua.
\par
Lastly, we note that because the volume-time-integral inequalities in (\ref{49'})--(\ref{52'}), (\ref{49''})--(\ref{52''}), and (\ref{52cor}) hold for all $V$, they hold as well with the volume integrals omitted (as long as all averages are done over macroscopic volumes), that is, with only the time integrals at each value of position $\vrp$.   In addition, as mentioned after (\ref{5W}), the materials to which (\ref{49'})--(\ref{52'}) and (\ref{49''})--(\ref{52''}) apply can contain conduction currents if the predominant energy {change} produced by the conduction current is in the form of heat (as is usually the case).  In other words, the usual conduction current $\vJ(\vrp,t)$ can be added to $\partial \vP(\vrp,t)/\partial t$ in (\ref{49'})--(\ref{52'}) and (\ref{49''})--(\ref{52''}).
\section{\lbl{BIAN}{APPLICATION OF ENERGY RELATIONS TO BIANISOTROPIC CONTINUA}}
In this section, the energy theorems summarized in the previous section are applied to linear,  passive, spatially nondispersive media to obtain frequency-domain expressions for internal energy densities in lossless media and inequalities that the linear constitutive relations must obey in lossless media. The most general linear, spatially nondispersive\footnote{\lbl{foot11}From Maxwell's equations, $\vH_\w$ is proportional to the spatial derivatives in $\nabla\times\vE_\w$ and $\vE_\w$ is proportional to the spatial derivatives in $\nabla\times\vH_\w$.  Thus, the bianisotropic constitutive relations in some contexts are considered inherently spatially dispersive.  Nonetheless, the curl spatial derivatives do not excite surface delta functions in $\vP_\w$ or $\vM_\w$.   Consequently, $\vnh\cdot(\vE_\w\times\vH_\w)$ is continuous across interfaces and our derived energy relations can be applied to the fields and polarizations that satisfy the bianisotropic constitutive relations in (\ref{Bi1}).} constitutive relations are those for bianisotropic media and are given in the frequency domain as
\begin{subequations}
\label{Bi1}
\be{Bi1a}
\vD_\w(\vrp) =\dep(\vrp) \cdot \vE_\w(\vrp) + \dtau(\vrp) \cdot \vH_\w(\vrp)
\ee
\be{Bi1b}
\vB_\w(\vrp) =\dmu(\vrp) \cdot \vH_\w(\vrp) + \dnu(\vrp) \cdot \vE_\w(\vrp)
\ee
\end{subequations}
where $\dmu(\vrp)$, $\dep(\vrp)$, and $[\dnu(\vrp), \dtau(\vrp)]$ are the permeability dyadic, the permittivity dyadic, and the magneto-electric dyadics, respectively.  Like the fields, they are, in  general,  functions of frequency $\w$ and position $\vrp$ within the media.
\par
First, assume the magnetic polarization of the bianisotropic material is {produced by paramagnetic dipoles}.  Then equations (\ref{M52}) and (\ref{52''}) apply, that is
\be{BiM52}
\int\limits_{t_0}^t \left[\frac{\partial\vD(\vrp,t')}{\partial t'}\cdot\vE(\vrp,t')
+ \frac{\partial\vB(\vrp,t')}{\partial t'}\cdot\vH(\vrp,t') \right] dt' \ge \frac{1}{2}\left[\eps_0 |\vE(\vrp,t)|^2 +\mu_0|\vH(\vrp,t)|^2\right]\,.
\ee
In \cite{Y&B, Yaghjian-Felsen}, the $\vE$ and $\vH$ in (\ref{BiM52}) were given a time dependence that began at a zero value and increased to a sinusoidal time dependence in order to prove the following inequalities for {paramagnetic} bianisotropic material that is lossless ($\dmu = \dmu^{\rm T*}$, $\dep = \dep^{\rm T*}$, $\dnu = \dtau^{\rm T*}$, with superscript ``T" denoting the transpose) in a finite frequency window about the frequency $\w$ of interest 
\be{Bi44}
\mbox{Re}\left\{\vE_\w^* \cdot (\w\dep)'\cdot  \vE_\w + \vH_\w^* \cdot (\w\dmu)'\cdot \vH_\w   +\vE_\w \cdot \left[(\w (\dnu^{\rm T} +\dtau^*)\right]'\cdot \vH_\w^*\right\} \ge \left[\epsilon_0 |\vE_\w|^2 + \mu_0|\vH_\w|^2 \right]  
\ee
\begin{subequations}
\label{Bi45}
\be{Bi45a}
[(\w \epsilon_{ll})' - \epsilon_0] \ge \w\epsilon_{ll}^{\prime}/2 \ge  0
\ee
\be{Bi45b}
[(\w \mu_{ll})' - \mu_0] \ge \w\mu^{\prime}_{ll}/2 \ge0
\ee
\end{subequations}
where the primes denote differentiation with respect to $\w$ and the double index $ll$ indicates the $xx$, $yy$, or $zz$ diagonal elements of the dyadic.
\par
If the magnetic polarization of the bianisotropic material is {produced by diamagnetic dipoles}, then equations (\ref{52}) and (\ref{52'}) apply, that is
\be{Bi52}
\int\limits_{t_0}^t \left[\frac{\partial\vD(\vrp,t')}{\partial t'}\cdot\vE(\vrp,t')
+ \frac{\partial\vB(\vrp,t')}{\partial t'}\cdot\vH(\vrp,t') \right] dt' \ge \frac{1}{2}\left[\eps_0 |\vE(\vrp,t)|^2 +\frac{1}{\mu_0}|\vB(\vrp,t)|^2\right]\,.
\ee
Using an analysis similar to the one in \cite{Y&B, Yaghjian-Felsen} but applied to (\ref{Bi52}) instead of (\ref{BiM52}), we find for a lossless frequency window
\be{Bi44D}
\mbox{Re}\left\{\vE_\w^* \cdot (\w\dep)'\cdot  \vE_\w + \vH_\w^* \cdot (\w\dmu)'\cdot \vH_\w   +\vE_\w \cdot \left[(\w (\dnu^{\rm T} +\dtau^*)\right]'\cdot \vH_\w^*\right\} \ge \left[\epsilon_0 |\vE_\w|^2 \!+\! \frac{1}{\mu_0}|\vB_\w|^2 \right]  
\ee
\begin{subequations}
\label{Bi45D}
\be{Bi45Da}
[(\w \epsilon_{ll})' - \epsilon_0] \ge \w\epsilon_{ll}^{\prime}/2 \ge  0
\ee
with a similar (though not identical) inequality for the diagonal elements of the permeability dyadic if the orientation of the $xyz$ coordinate system is chosen at each point in space to make the permeability dyadic diagonal (which is always possible because the lossless permeability dyadic is Hermitian), namely
\be{Bi45Db}
[(\w \mu_{ll})' - \mu^2_{ll}/ \mu_0] \ge \w\mu^{\prime}_{ll}/2 \ge0\,.
\ee
\end{subequations}
Note that as $\w\to0$,  (\ref{Bi45a}) and (\ref{Bi45Da}) imply that for both {paramagnetic} and diamagnetic material
\be{Bi2}
\eps_{ll}(\w\to0)-\eps_0 \ge 0
\ee
(assuming $\lim_{\w\to0}(\w\eps^{\prime}_{ll})=0$).
Also, (\ref{Bi45b}) implies that
\be{Bi3}
\mu_{ll}(\w\to0)-\mu_0 \ge 0
\ee
for {paramagnetic} material (assuming $\lim_{\w\to0}(\w\mu^{\prime}_{ll})=0$).  However, (\ref{Bi45Db}) implies that
\be{Bi4}
\mu_{ll}(\w\to0)[1-\mu_{ll}(\w\to0)/\mu_0] \ge 0
\ee
or, equivalently
\be{Bi5}
0\le\mu_{ll}(\w\to0)\le \mu_0
\ee
for diamagnetic material (assuming $\lim_{\w\to0}(\w\mu^{\prime}_{ll})=0$).
\subsection{\label{Subtleties}{Subtleties associated with diamagnetic continua}}
The inequalities in (\ref{Bi45a}) and (\ref{Bi45b}) can also be obtained from the Kramers-Kronig causality relations for $\dep$ and $\dmu$, respectively \cite[sec. 84]{LLP}. Thus, one can ask why (\ref{Bi45b}) does not hold also for diamagnetic material or for an array of diamagnetic inclusions that behaves as a continuum.  This question was answered in detail in \cite{YAS1,YAS2} with the answer summarized in the Conclusion of \cite{YAS2} as,\\\\
``At high enough frequencies $\w$, every natural material or artificial material (metamaterial) no longer behaves as a continuum satisfying the traditional time-harmonic dipolar macroscopic Maxwell equations with spatially nondispersive constitutive parameters.  Although this departure from a continuum behavior at high frequencies can often be ignored with impunity for the electric and para/ferro(i)magnetic polarization of materials and metamaterials, we show in the Introduction that it is mathematically impossible to characterize a material or metamaterial that is diamagnetic and lossless at low frequencies $\w$ by a causal [obeying the Kramers-Kronig relations] spatially nondispersive permeability that satisfies continuum passivity conditions and whose value approaches the permeability of free space [$\mu_0$] as the frequency $\w$ approaches infinity.  Moreover, this noncausality in the spatially nondispersive dipolar continuum description of diamagnetism is more fundamental than the noncausality, discussed in \cite{AYSS}, introduced by the point dipole approximation for scattering from the inclusions (molecules) and is not removed by including higher-order multipole moments in the spatially nondispersive continuum formulation of Maxwell's equations."\\\\
It is possible for diamagnetic causality to be restored and for (\ref{Bi45b}) to remain valid for a lossless diamagnetic continuum if $\mu_0$ is changed to $\mu_{ll}^\infty=\mu_{ll}(\w\to\infty)<\mu_0$ such that $\mu_{ll}(\w\to0)-\mu_{ll}^\infty \ge0$.  However, this possibility is unrealistic for spatially nondispersive continua because the magnetic dipole moment of a given molecule or inclusion excited by a free-space incident field generally approaches zero as $\w\to\infty$ and thus $\mu_{ll}^\infty = \mu_0$ \cite{YAS1,YAS2}.\footnote{\lbl{foot12}If one postulates a spatially nondispersive linear continuum with a hypothetical causal diamagnetic $\dmu$ that approaches $\mu^\infty\dI =\mu(\w\to\infty)\dI$, where causality requires that $\mu_{ll}(\w\to0)\ge\mu^\infty$, then the method
used by Glasgow et al. \cite{Glasgow}
for continua with causal, re\-ci\-procal, pas\-sive $\dep$ and implicitly {paramagnetic} $\dmu$ that approaches $\mu_0\dI$ as $\w\to\infty$ can be applied to the hypothetical causal, reciprocal, passive diamagnetic $\dmu$ that approaches $\mu^\infty\dI$ as $\w\to\infty$ to obtain the diamagnetic energy inequality
\bea{causaldiam}
-\int\limits_S\int\limits_{t_0}^t \vnh\cdot[\vE(\vrp,t')\times\vH(\vrp,t')]dt' dS=\int\limits_V\int\limits_{t_0}^t \left[\frac{\partial\vD(\vrp,t')}{\partial t'}\cdot\vE(\vrp,t')
+ \frac{\partial\vB(\vrp,t')}{\partial t'}\cdot\vH(\vrp,t') \right] dt'dV\nonumber\\ \ge \frac{1}{2}\int\limits_V\left[\eps_0 |\vE(\vrp,t)|^2 +\mu^\infty|\vH(\vrp,t)|^2\right]dV \hspace{1.5cm}
\eea
from which one finds instead of (\ref{Bi5}) for zero loss as $\w\to0$
\be{causaldiamBi5}
\mu_{ll}(\w\to0)\ge \mu^\infty
\ee
a result that merely repeats the requirement for a causal diamagnetic-continuum permeability.  Moreover,  neither natural materials nor metamaterials are characterized by spatially nondispersive constitutive parameters above frequencies for which the average electrical separation distance of the molecules or inclusions becomes a significant portion of a wavelength and, thus, one does not generally have a causal expression for diamagnetic material that reveals a reliable analytic high-frequency continuation (in particular, a $\mu^\infty<\mu_0$) that can be used in (\ref{causaldiam})--(\ref{causaldiamBi5}).  The inequalities in (\ref{49'})--(\ref{52'}) for electric-dipolar-diamagnetic macroscopic continua and (\ref{49''})--(\ref{52''}) for electric-dipolar-{paramagnetic} macroscopic continua are derived without requiring linearity or reference to any particular constitutive relations or parameters.} 
\par
Another subtlety arises with regard to a lossless (in a frequency window), homogeneous, spatially nondispersive, bianisotropic continuum.  In such a continuum, it can be proven \cite{YMM} that the group velocity $\vv_g(\vk_{\rm o})$ of a source-free wave packet (with propagation vectors $\vk$ concentrated about the vector $\vk_{\rm o}$ and a narrow-band frequency spectrum $\w(\vk)$ such that the envelope of the wave packet is a virtually unchanging function $\vg[\vrp-\nabla\!_k \w(\vk_{\rm o})t]$) equals the energy-transport velocity $\vv_e(\vk_{\rm o})$; specifically
\be{Bi6}
\vv_g(\vk_{\rm o})\equiv \nabla\!_k \w(\vk_{\rm o}) = \frac{\vS(\vk_{\rm o})}{U(\vk_{\rm o})}\equiv \vv_e(\vk_{\rm o})
\ee
where
\begin{subequations}
\label{Bi7}
\be{Bi7a}
 \vS(\vk_{\rm o})=\frac{1}{2}{\rm Re}[\vE_{\rm o}\times\vH^*_{\rm o}]
\ee
and
\be{Bi7b}
U(\vk_{\rm o})=\frac{1}{4}\mbox{Re}\left\{\vE_{\rm o}^* \cdot (\w\dep)'\cdot  \vE_{\rm o} + \vH_{\rm o}^* \cdot (\w\dmu)'\cdot \vH_{\rm o}   +\vE_{\rm o} \cdot \left[(\w (\dnu^{\rm T} +\dtau^*)\right]'\cdot \vH_{\rm o}^*\right\}\,.
\ee
\end{subequations}
The subscripts ``o" on $\vE_{\rm o}$ and $\vH_{\rm o}$ refer to these fields evaluated at the frequency $\w(\vk_{\rm o})$.  For {paramagnetic} continua, the inequality in (\ref{Bi44}) shows from (\ref{Bi6}) that
\be{Bi8}
|\vv_g(\vk_{\rm o})|=|\vv_e(\vk_{\rm o})|\le 2\frac{|{\rm Re}[\vE_{\rm o}\times\vH^*_{\rm o}]|}{\eps_0|\vE_{\rm o}|^2 
+ \mu_0|\vH_{\rm o}|^2}\,.
\ee
Since 
\bea{Bi9}
|{\rm Re}[\vE_{\rm o}\times\vH^*_{\rm o}]|\le |\vE_{\rm o}||\vH_{\rm o}|=\frac{c}{2}[\eps_0|\vE_{\rm o}|^2 
+ \mu_0|\vH_{\rm o}|^2 -(\sqrt{\eps_0}|\vE_{\rm o}|-\sqrt{\mu_0}|\vH_{\rm o}|)^2]\nonumber\\\le \frac{c}{2}[\eps_0|\vE_{\rm o}|^2 + \mu_0|\vH_{\rm o}|^2]
\eea
where the free-space speed of light is given by $c=1/\sqrt{\mu_0\eps_0}\,$, this inequality in (\ref{Bi9}) reduces (\ref{Bi8}) to
\be{Bi10}
|\vv_g(\vk_{\rm o})|=|\vv_e(\vk_{\rm o})|\le c
\ee
confirming that the speed of a lossless wave packet in the {paramagnetic} continuum cannot be faster than the speed of light in free space.
\par
The subtlety arises by repeating the derivation for a diamagnetic continuum wherein the inequality in  (\ref{Bi44}) must be replaced by the inequality in (\ref{Bi44D}), so that (\ref{Bi8}) is replaced by
\be{Bi11}
|\vv_g(\vk_{\rm o})|=|\vv_e(\vk_{\rm o})|\le 2\frac{|{\rm Re}[\vE_{\rm o}\times\vH^*_{\rm o}]|}{\eps_0|\vE_{\rm o}|^2 
+ |\vB_{\rm o}|^2/\mu_0}
\ee
which can be rewritten from (\ref{Bi9}) as
\be{Bi12}
|\vv_g(\vk_{\rm o})|=|\vv_e(\vk_{\rm o})|\le c\frac{\eps_0|\vE_{\rm o}|^2 + \mu_0|\vH_{\rm o}|^2}{\eps_0|\vE_{\rm o}|^2 
+ |\vB_{\rm o}|^2/\mu_0}\,.
\ee
Considering, for example, a lossless isotropic diamagnetic continuum at low frequencies such that $\vB_{\rm o}=\mu\vH_{\rm o}$, $0<\mu<\mu_0$, we see that the right-hand side of (\ref{Bi12}) can be greater than the free-space speed of light $c$.  This result does not actually imply that the speed of the wave packet can be greater than the free-space speed of light $c$ in diamagnetic materials, but it is disappointing that the right-hand side of (\ref{Bi12}) does not come out to be equal to $c$ as it does for {paramagnetic} materials.  What it does imply, however, is that the equality in (\ref{Bi6}) between the group and energy-transport velocities in lossless, linear, spatially nondispersive continua does not provide a robust method for proving that the magnitude of these velocities are less than or equal to the free-space speed of light.  
\par
The fundamental principle that can be invoked to obtain this result is the Einstein mass-energy relation.  Since a fixed amount of energy (call it $W_0$) in the source-free wave packet of width $\ell$ travels a distance $L\gg \ell$ in the lossless continua without significantly changing shape and with velocity $\vv$ equal to the group and energy-transport velocities ($\vv = \vv_g=\vv_e$), the effective moving and rest masses, $m$ and $m_0$, of this wave packet of energy $W_0$ are found from the Einstein mass-energy relation as
\be{Bi13}
W_0 = mc^2= \frac{m_0c^2}{\sqrt{1-|\vv|^2/c^2}}
\ee
which implies 
\be{Bi14}
|\vv|= |\vv_g| = |\vv_e| \le c 
\ee
for diamagnetic as well as {paramagnetic} continua.  In other words, if a passive medium can propagate a fixed-shape bundle of a fixed amount of energy, then the speed of propagation of this bundle cannot be greater than the free-space speed of light.  A lossless bianisotropic continuum supports just such a fixed-energy, distortionless wave packet.  Whereas in a lossy, highly dispersive continuum, both the amount of energy and the local shape of the wave packet envelope can change and this distortion can make parts of the envelope (but not the leading edge of the wave packet) travel faster than the free-space speed of light \cite{superluminal}.
\par
One of the simplest applications of the inequality in (\ref{Bi14}) is to a lossless isotropic magnetodielectric material at low frequencies where its $\eps$ and $\mu$ are virtually independent of frequency.  Then the group, energy-transport, and phase velocities of a low-frequency pulse in the material are approximately the same and equal to $1/\sqrt{\mu\eps}$.  Thus, (\ref{Bi14}) reveals that
\be{Bi15}
\frac{\eps}{\eps_0} \ge \frac{\mu_0}{\mu}\,.
\ee
For {paramagnetic} material, $\mu >\mu_0$ and the inequality in (\ref{Bi15}) gives less information than the one in (\ref{Bi2}) and thus it can be ignored.  However, for diamagnetic material, (\ref{Bi15}) can be expressed as 
\be{Bi16}
\eps \ge \mu_{\rm r}^{-1} \eps_0
\ee
where $\mu_{\rm r}^{-1} = \mu_0/\mu$ is greater than $1$ and thus (\ref{Bi16}) replaces (\ref{Bi2}).  Wood and Pendry \cite{W&P} and Sohl et al. \cite{SGB} have also obtained the inequality in (\ref{Bi16}) for diamagnetic materials.
\section{{CONCLUSION}}
With the insight obtained from a relatively simple, yet rigorous proof that averaging the microscopic fields of {classical} discrete dipoles over macroscopic volumes leads to the same Maxwell equations and cavity fields obtained for mathematically defined fields in ideal continuous dipolar media, positive semi-definite expressions are derived for macroscopic time-domain energy density in passive, spatially nondispersive (more generally, continuous $\vnh\cdot\vE\times\vH$ across interfaces) dipolar continua.  The derivation proceeds from the underlying microscopic {Maxwell} equations satisfied by the microscopic fields of the electric charge and current that {produce} the distribution of discrete, bound, dipolar classical models of the molecules or inclusions comprising material or metamaterial continua.   The bound microscopic electric dipoles are assumed to have either zero dipole moments before any external fields are applied or randomly oriented, fixed-magnitude dipole moments that can be aligned by the external fields.  In both cases, the applied fields can extract no energy from the initial electric dipoles and thus the electric dipoles are ``unconditionally passive".
\par
The microscopic derivation reveals two distinct positive semi-definite (nonnegative) macroscopic time-domain energy expressions, one that applies to diamagnetic continua and another that applies to paramagnetic continua, which includes ferro(i)magnetic  and antiferromagnetic materials.  The diamagnetic dipoles are unconditionally passive because their Amperian magnetic dipole moments are zero in the absence of applied fields.  The analysis of the paramagnetic continua, which are defined in terms of magnetization caused by the alignment of randomly oriented permanent (yet slightly changeable in magnitude as they rotate in an external magnetic field) Amperian magnetic dipole moments that dominate any induced diamagnetic magnetization, is greatly simplified by first proving that the microscopic power equations for rotating ``permanent" Amperian magnetic dipoles (which are not unconditionally passive because energy can be extracted from their initial magnetic dipole moments)  reduce effectively to the same power equations obeyed by rotating unconditionally passive magnetic-charge magnetic dipoles.
\par
The difference between the paramagnetic and diamagnetic energy expressions is equal to a ``hidden energy" that parallels the hidden momentum often attributed to Amperian magnetic dipoles.  It is noteworthy that the microscopic derivation reveals that this hidden energy is supplied by the change of inductive energy in the Amperian magnetic dipole moments as the ``permanent" dipoles align in an applied field, even though the magnitudes of the induced magnetic dipole moments are negligible compared to the magnitudes of the total magnetic dipole moments.
\par
The macroscopic, positive semi-definite, time-domain energy expressions are applied to lossless bianisotropic media to determine the inequalities obeyed by the frequency-domain bianisotropic constitutive parameters, namely the permittivity, permeability, and magneto-electric dyadics.  As one would expect, the inequalities obeyed by diamagnetic and {paramagnetic} permeabilities of bianisotropic media are appreciably different.  Subtleties associated with the causality as well as the group and energy-transport velocities for diamagnetic media are discussed in view of the diamagnetic inequalities.  In particular, it is shown that general proofs of the group and energy-transport speeds being less than (or equal to) the free-space speed of light in lossless media must rely ultimately on the Einstein mass-energy relation.
\par
Lastly, it should be noted that materials or metamaterials have not been considered in which their molecules or inclusions consist of a combination of induced diamagnetic dipoles (zero initial Amperian magnetic dipole moments) and paramagnetic dipoles (predominant ``permanent" magnetic dipole moments) such that both the diamagnetism and paramagnetism are comparable.  A metamaterial array with inclusions containing ferromagnetic material and electric conductors (and/or dielectrics) creating diamagnetism would be one such possibility.  Then, neither the diamagnetic nor paramagnetic energy expression derived in this paper would necessarily apply to this combined magnetization.  The generalization of positive semi-definite macroscopic energy expressions to materials with comparable combined paramagnetic and diamagnetic magnetization is a subject for future research.
%
\ack
This research was supported under the U.S. Air Force Office of Scientific Research (AFOSR) Grant \# FA9550-16-C-0017 through Dr. A. Nachman. Helpful discussions with Professors M.G. Silveirinha and A. Al\`u are gratefully acknowledged.
%
%
%
%

%
%
\end{document}